\documentclass[preprint2]{aastex}

\newcommand{\mic}{$\mu$m}
\newcommand{\ebv}{E(B$-$V)}
\newcommand{\kms}{km s$^{-1}$}

\newcommand{\smass}{$10^6$ $M_\odot$}
\newcommand{\mass}{$M_\odot$}
\newcommand{\massyr}{$M_\odot$ yr$^{-1}$}
\newcommand{\ha}{H$\alpha$}
\newcommand{\ewha}{EW(H$\alpha$)}
\newcommand\colzero {\null}
\newcommand\cola {&}
\newcommand\colb {&}
\newcommand\colc {&}
\newcommand\cold {&}
\newcommand\cole {&}
\newcommand\colf {&}
\newcommand\colg {&}
\newcommand\colh {&}
\newcommand\coli {&}
\newcommand\colj {&}
\newcommand\colk {&}

\newcommand\eol{\\}

\shorttitle{Arp 82}
\shortauthors{Hancock {\it et al.}}

\begin{document}

\title{Large-scale Star Formation Triggering in the Low-mass Arp 82 System: A Nearby Example of Galaxy Downsizing Based on UV/Optical/Mid-IR Imaging}

\author{Mark Hancock\altaffilmark{1}, Beverly J. Smith\altaffilmark{1}, Curtis Struck\altaffilmark{2}, Mark L. Giroux\altaffilmark{1}, Philip N. Appleton\altaffilmark{3}, Vassilis Charmandaris\altaffilmark{4}$^{,}$\altaffilmark{5}$^{,}$\altaffilmark{6}, \& William T. Reach\altaffilmark{3}}
\email{hancockm@etsu.edu, smithbj@etsu.edu, curt@iastate.edu, girouxm@etsu.edu, apple@ipac.caltech.edu, vassilis@physics.uoc.gr, \& reach@ipac.caltech.edu}

\altaffiltext{1}{Department of Physics, Astronomy, and Geology, East Tennessee State University, Box 70652, Johnson City, TN 37614}
\altaffiltext{2}{Department of Physics and Astronomy, Iowa State University, Ames IA 50011}
\altaffiltext{3}{Spitzer Science Center, California Institute of Technology, Pasadena CA 91125}
\altaffiltext{4}{Department of Physics, University of Crete, GR-71003, Heraklion Greece}
\altaffiltext{5}{Chercheur Associ\'e, Observatoire de Paris, 
F-75014,  Paris, France}
\altaffiltext{6}{IESL / Foundation for Research and Technology- Hellas, PO Box 1527, 71110, Heraklion, Greece}

\begin{abstract}

As part of our Spitzer Spirals, Bridges, and Tails project to help
understand the effects of galaxy interactions on star formation, we
analyze GALEX ultraviolet, SARA optical, and Spitzer infrared images
of the interacting galaxy pair Arp 82 (NGC 2535/6) and compare to a
numerical simulation of the interaction.   We investigate the
multi-wavelength properties of several individual star forming
complexes (clumps).   Using optical and UV colors, \ewha, and
population synthesis models we constrain the ages of the clumps and
find that the median clump age is $\sim$9 Myr.   The clumps have
masses ranging from a few $\times10^6$ to $10^9$ \mass.  In general,
the clumps in the tidal features have similar ages to those in the
spiral region, but are less massive.  The clumps provide 33\%, 36\%,
and 70\% of the FUV, 8.0 \mic, and 24 \mic\ emission, respectively.
The 8 \mic\ and 24 \mic\ luminosities are used to estimate the
far-infrared luminosities and the star formation rates of the clumps.
The total clump star formation rate is $\sim$2.0$\pm0.8$ \massyr,
while the entire Arp 82 system is forming stars at a rate of
$\sim$4.9$\pm2.0$ \massyr.  We find, for the first time, stars in the
H~{\scriptsize I} arc to the southeast of  the NGC 2535 disk.
Population synthesis models indicate that all of the observed
populations have young to intermediate ages.  We conclude that
although the gas disks and some old stars may have formed early-on,
the progenitors may have been late-type or low surface brightness and
the evolution of these galaxies seems to have halted until the recent
encounter.

\end{abstract}

\keywords{galaxies: starbursts --- galaxies: interacting --- galaxies: numerical models --- galaxies: individual (Arp 82)}

\section{INTRODUCTION}

Studies of broad band optical and near infrared colors gave early
indications that galaxy interactions induce star formation (e.g.,
\citealp{lar78,str78}).  There is now a wealth of observational and
modeling information on how interactions and mergers can drastically
modify the morphology and star formation rates of galaxies (see review
by \citealp{str99}).  The Infrared Astronomical Satellite (IRAS)
revealed a new population of high infrared luminosity galaxies with
large star formation rates (\citealp{soi87,smi87}) that are the result
of major mergers between gas-rich progenitors \citep{san88}.  In
several interacting and merging galaxies, individual knots of star
formation have been observed (e.g. \citealp{meu95,whi03}).  Many such
knots are likely to be the progenitors of globular  clusters
(e.g. \citealp{han03,wei04}).  Another type of star forming clump
associated with interacting galaxies are Tidal Dwarf Galaxies (TDG)
(\citealp{mir92,duc94}).  These TDGs are frequently formed in tidal
tails during the mergers of dusty, gas rich galaxies (see for example,
\citealp{too72,san96}).

As part of our Spitzer Cycle 1 `Spirals, Bridges, and Tails' (SB\&T)
project, we are investigating whether or not interacting but not yet
merging galaxies have heightened star formation properties. In the
SB\&T survey \citep{smi06}, we obtained 3.6$-$24 \mic\ images of 35
nearby interacting systems selected from the  Arp Atlas of Peculiar
Galaxies (1966).  We previously  presented a detailed study of one of
these galaxies, Arp 107, in \citet{smi05}.  In the current paper we
investigate a second SB\&T system,  the interacting pair Arp 82 (NGC
2535/6).  We have obtained UV, visible, and mid-IR images of Arp 82 from
the Galaxy Evolution Explorer (GALEX), Southeastern Association for
Research in Astronomy (SARA), and Spitzer telescopes respectively.
Similar multi-wavelength studies of other interacting galaxies have
been done by \citet{cal05} and \citet{elm06}.

With these high detail mid-IR maps, we can study star formation
processes at wavelengths where extinction is significantly reduced
compared with studies at UV or optical wavelengths.
Observations in the ultraviolet are ideal for investigating the young
star forming regions since in this wavelength range hot young stars
dominate the spectral energy distribution and can be easily
distinguished from older populations.  Interacting galaxies are
particularly good targets for such studies, since comparisons to
dynamical models of the interactions can provide information about
time scales, and give clues to star formation triggering mechanisms
(e.g., \citealp{str03}).

Arp 82 is more quiescent than the median Arp galaxies of the
\citet{bus88} sample, with a relatively low 60 \mic/100 \mic\ flux
ratio of 0.32 and a total far-infrared luminosity of
$\sim$1.6$\times10^{10}$ L$_{\odot}$.  The total H~{\scriptsize I}
mass of Arp 82 is  $2.3\times10^{10}$ \mass, of which
$2.3\times10^{9}$ \mass\ is associated  with NGC 2536 (the smaller
companion) \citep{kau97}, typical  of Low Surface Brightness (LSB)
galaxies in the sample of \citet{one04}.  The total 3.6 \mic\
luminosity is low compared to `normal' spirals, thus  Arp 82 is a
relatively low mass system (see \citealp{smi06}).  Studies of such
systems are relevant to the question of whether most of the star
formation since z$=$1 is occurring in low-mass systems (e.g., the so
called downsizing idea, first suggested by \citealp{cow96}).

Based on optical spectroscopy, NGC 2535 and 2536 have been classified
as both H~{\scriptsize II} galaxies \citep{kee85} or as a Low
Ionization Emission Line Region (LINER) and an H~{\scriptsize II}
galaxy respectively \citep{dah85}.  NGC 2535,  has a bright oval of
star formation shaped like an eyelid.  This type of ocular structure
has been observed  in other galaxies, e.g. IC 2163 and NGC 2207
\citep{elm06}, and is presumably the result of large-scale gaseous
shocks from a grazing prograde encounter.  There is an interesting
tidal `arc' in the southeast seen clearly in H~{\scriptsize I} maps
\citep{kau97}.

We adopt a distance of 57 Mpc for Arp 82.  This was calculated using
velocities from \citet{hay97} for NGC 2535 and \citet{fal99} for NGC
2536, a Hubble constant of 75 \kms\ Mpc$^{-1}$, and the \citet{sche80}
Virgo-centric in-fall model with parameters as in \citet{hec98}.  The
projected separation between the two galaxies in Arp 82 is 102\arcsec\
(28 kpc).

The present paper is organized as follows.  In \S2 we describe our
multi-wavelength observations and outline the data reductions.  We
describe our analysis in \S3 and discuss our findings in \S4.  In \S5
we describe our numerical collision model.  We compare Arp 82 to other
systems in \S6 and finally, we summarize the paper in \S7.

\section{OBSERVATIONS AND DATA REDUCTIONS}

\subsection{UV Observations}

We obtained ultraviolet images of Arp 82 on February 13 and 24, 2005
using the Galaxy Evolution Explorer (GALEX) satellite \citep{mar05}.
The galaxy was imaged in both the far-UV (FUV) and near-UV (NUV) bands
covering the wavelength ranges 1350\AA\ to 1750\AA\ and 1750\AA\ to
2800\AA, respectively.  The total integration times in the  FUV and
NUV filters were 1684 and 3019 seconds, respectively.  GALEX uses two
65 millimeter diameter, micro-channel plate detectors producing
circular images of the sky with 1.2 degree diameter and 5\arcsec\
resolution in the two UV bands.  The GALEX images were reduced and
calibrated through the GALEX pipeline.  The details
of these and our other observations  are listed in Table 1.

\subsection{Optical Observations}

Optical images of Arp 82 were obtained on December 18, 2004 under
clear skies using the Southeastern Association for Research in
Astronomy (SARA) 0.9 meter telescope on Kitt Peak.  The observations
were made with a 2048 $\times$ 2048 Axiom Apogee CCD, with the binning
set to 2 $\times$ 2.  This gives a pixel size of 0\farcs52 and a field
of view of 8\farcm9 $\times$ 8\farcm9.  A narrow band filter centered
at 664 nm (FWHM $\approx$ 70\AA) was used to measure the redshifted
\ha\ emission; this filter also contained the $\lambda$6583\AA\
[N~{\scriptsize II}] line.  A broadband R filter was used to measure
the continuum.  In the 664 nm and R filters, respectively, 16 and 5
exposures of 10 minutes each were obtained, along with sky flats, bias
and dark frames.  The white dwarf stars HZ 14 and Hiltner 600 were
also observed for calibration purposes.  The seeing was
$\approx$2\farcs5.  The details of the observations are listed in
Table 1.

The optical data were reduced in a standard way using the Image
Reduction and Analysis Facility (IRAF\footnote{IRAF is distributed by
the National Optical Astronomy Observatories, which are operated by
the Association of Universities for Research in Astronomy, Inc., under
cooperative agreement with the National Science Foundation.})
software.  Continuum subtraction was accomplished using a scaled R
band image.  The calibration conversions derived from the two standard
stars agreed within 21\%.   The larger galaxy NGC 2535 contributes
81\% of the total \ha+[N~{\scriptsize II}] emission.  Our \ha\ fluxes
are a factor of $\approx$2.5 less than the \citet{ken87} values.  We
are unable to explain why our H$\alpha$ fluxes are less than that of
\citet{ken87} for the same system. We note that had we adopted the
\citet{ken87} fluxes, Arp 82 would lie even further from the median
relationship of L(FIR)/L(H$\alpha$) for interacting galaxies found by 
\citet{bus88}.

\subsection{IR Observations}
Arp 82 was imaged with the ``InfraRed Array Camera'' (IRAC,
\citealp{faz04}) on the Spitzer Space Telescope \citep{wer04}
at 3.6, 4.5, 5.8, and 8.0 \mic\ on November 1, 2004, as
well as with the ``Multi-band Imager and Photometer for Spitzer''
(MIPS, \citealp{rie04}) at 24 \mic\ on April 2, 2005.  The IRAC detectors
consist of two 256$\times$256 square pixel arrays with a pixel size
of 1\farcs22 resulting in a total field of view of
5\farcm2$\times$5\farcm2.  The full-width at half maximum (FWHM) of
the point spread function (PSF) varies between 1\farcs5 and
2\arcsec \, across the different IRAC channels.  The MIPS detector
\citep{rie04} at 24 \mic\ uses a 2\farcs45 pixel size array of
128$\times$128 elements also resulting in a field of view of
5\farcm2$\times$5\farcm2. The image at this wavelength is
characterized by a PSF with a FWHM of $\sim$6\arcsec.

Each IRAC observation was performed using a sequence of six 12 second
frames in a dithered cyclic pattern. The total on source time for each
filter was 72 seconds.  At 24 \mic\ a similar mapping technique of a
series of 10 second frames led to a total on source coverage of 321
seconds.  The data were reduced with standard procedures (i.e.,
dark-current subtraction, cosmic ray removal, non-linearity
correction, flat-fielding and mosaicing) using pipeline version
13.2 of the Spitzer Science Center\footnote{see
http://ssc.spitzer.caltech.edu/postbcd/}.  The absolute pointing
accuracy of Spitzer is better than $\sim$1\arcsec\ and the 1$\sigma$
relative astrometric uncertainty is less than $\sim$0\farcs3 in the
IRAC and MIPS data.  The details of the observations are listed in
Table 1.

\section{DATA ANALYSIS}

\subsection{Morphology and Identifying the Clumps}

In Figure 1 we present the FUV, NUV, \ha, R band, 3.6 \mic, 4.5 \mic,
5.8 \mic, 8.0 \mic, and 24 \mic\ images of Arp 82 respectively.   The
bridge and countertail are easily seen in the FUV and NUV images.  The
tail region is much more prominent in the UV than in the IR as one
would expect.  The UV traces older stars than \ha, which is a tracer
of massive stars.  Since \ha\ correlates with warm dust in emission
(\citealp{rou01} and references therein), we expect the 8 \mic\ to be
very close to areas harboring photon dominated regions (PDRS) and
massive stars.

Several bright star forming regions (clumps) can be seen in the 8
\mic\ image (Figure 2).   We have identified 26 clumps by visual
inspection of the 8 \mic\ image.  To strengthen our detection
criterion, we required the measured fluxes of the clumps to be greater
than 3$\sigma$ of the sky background times the square root of the
number of pixels in the photometric aperture.  All 26 clumps were
detected in the FUV, NUV, R band, 8 \mic, and 24 \mic\ images,  24
were detected at the 3$\sigma$ level in H$\alpha$, 25 were detected at
3.6 \mic, 23 were detected at 4.5 \mic, and 24 were detected at 5.8
\mic.

Clump 13, in the very center of
NGC 2535, is clearly visible in the IRAC and MIPS images.  It is also
visible in the \ha, R, and NUV images, however, it almost disappears
in FUV.  This could be the result of reddening.  The L(IR)/L(FUV)
ratio (see \S 4.2) is a measure of the total UV opacity.  L(IR)/L(FUV)
is greatest for knot 13, consistent with a large reddening.  The
bright clump at the beginning of the northern tail, 21, is among the
brightest of the star forming regions in Arp 82 at 8 \mic\ and 24
\mic.  The bright clump at the end of the northern tail, 22, has a
significantly higher H$\alpha$/mid-infrared ratio than the other
clumps.

Clumps 12, 14, and 15 make up the bright `ocular ring' at the
outer edges of the spiral region in NGC 2535.  Notice that the
`ocular ring' is bright in all the bands.  Dramatic  `beads on a
string' features are visible: chains of evenly-spaced star formation
complexes (Figure 1).   These clumps are separated by a characteristic
distance scale of $\sim$1 kpc, implying that they are caused by
gravitational collapse of interstellar gas clouds under self-gravity
\citep{elm96}.

In addition to the 26 clumps described above, we identified 4 clumps
along the H~{\scriptsize I} arc to the southeast by visual inspection
of the FUV image.  These additional clumps, 27, 28, 29, and 30, can
also be seen clearly in H~{\scriptsize I} \citep{kau97}.  See \S 4.5
for details on these additional clumps, which are discussed separately.

\subsection{Background Subtraction}

To properly subtract an appropriate sky value from our flux
measurements, to minimize contributions from neighboring clumps, to
correct for a gradient in the 5.8 \mic\ image, and to remove the
effect of muxbleed in the 8.0 \mic\ image,  we created and subtracted
background images.  These background images were created by masking
the clumps in the original images and replacing them with the median
background values.  Then the entire background images (including the
non clump component of the parent galaxies) were boxcar smoothed.  
The contamination by the parent galaxies was therefore removed.  We
determined the replacement values by using the IRAF task IMEXAM with a
50$\times$50 pixel window to measure the median sky values in several
regions near the galaxies.  The sizes of the replacement window were
chosen to be bigger than our photometric apertures and the smoothing
windows were chosen to be much larger than the replacement windows.
The IRAF task BOXCAR  was used to smooth the background images.
Finally the smoothed background images were subtracted from the
original images.  Photometry was performed on the background
subtracted images.

\subsection{Photometry}

The measured fluxes of the clumps  are listed in Table 2.  The first
column is the clump ID, columns 2-3 are the RA and DEC, columns 4-5
are the UV fluxes, columns 6-7 are the optical fluxes, and columns
8-12 are the IR fluxes.  The measured colors are listed in Table 3.

The photometry was performed on the FUV and NUV images with the IRAF
task PHOT in the DAOPHOT package.  We used circular apertures of 3
pixel radii (4\farcs5) centered on the positions of the clumps.
Using stars in the field, we determined an FUV aperture
correction of $\sim$1.15 and an NUV aperture correction of $\sim$1.45,
by comparing fluxes in 3 pixel (4\farcs5) with those in 25 pixel
(37\farcs5) apertures.   The UV fluxes were converted to magnitudes on
the AB system \citep{oke90}.

The H$\alpha$ and R band photometry was performed on the SARA images
using 9 pixel (4\farcs5) radii apertures.  As discussed in \S 3.2, we
subtracted  a background image.  Calibration was accomplished using
the standard  stars HZ 14 and Hiltner 600 \citep{oke74}, and
correcting the \ha\ flux for [N~{\scriptsize II}] by assuming
\ha/[N~{\scriptsize II}]$\sim$3, typical of H~{\scriptsize II} regions
\citep{ost89}.   We determined the aperture correction for the SARA
\ha\  and R band  images by measuring the flux of several stars on the
images through 7.5 pixel and 25 pixel apertures.  The SARA \ha\
aperture correction is $\sim1.18$ and the R band aperture correction
is $\sim$1.33.

The photometry on all 4 IRAC images was done with 3 pixel (3\farcs6)
apertures.   As with the UV and optical data, no sky annuli were used
due to crowding.  Instead  we created and subtracted a background
image as described previously.  We used aperture corrections of 1.12
for the 3.6 \mic\ and 4.5 \mic\  bands, 1.14 for the 5.8 \mic\ band,
and 1.23 for the 8.0 \mic\ band (IRAC Data Handbook v2.0).

The 24 \mic\ fluxes of the clumps were measured with  a 1.22 pixel
(3\arcsec) aperture.  The same technique was used to  remove
background as described above.  According to the MIPS Data Handbook, the
aperture correction for a 3\arcsec aperture is 3.097.

We measured the flux of the entire Arp 82 system in all
our observed wavebands using the IRAF task IMSTAT on rectangular
regions covering the full observed extent.  The sky background was
determined from the mean counts in multiple box regions located around
Arp 82,  selected to avoid bright stars.  

To determine the uncertainties in the flux measurements, we assumed that 
Poisson noise was the dominant component.  To account for the effects of
different methods of sky subtraction, we repeated the photometry on the
original galaxy images in each band using the mode of a sky annulus of
radii 1 pixel larger than the photometric aperture radii and width 5
pixels as our sky subtraction.  We took the square root of the
difference in counts between the annulus subtracted measurements and
the background subtracted measurements as an additional source of
uncertainty.  This additional uncertainty was added in quadrature to
the statistical uncertainties.  Typically the additional uncertainties
were less than 10\%, except in the crowded regions where the
additional uncertainties were typically 15\%-20\%.

\section{RESULTS}

\subsection{Spitzer Colors and Gradients}

To study the IR color variations of the clumps with position, we
plotted the various IR colors as a function of distance from the NGC
2536 nucleus.   Figure 3 plots the [3.6]$-$[4.5] colors as a function
of distance  from NGC 2536.  Open squares represent clumps in NGC
2536, x's represent clumps in the bridge, stars represent clumps in
the spiral and filled squares represent clumps in the countertail.
Figure 4 plots the [4.5]$-$[5.8]  colors against
distance from NGC 2536.   Figure 5 shows the [5.8]$-$[8.0]
colors as a function of distance from NGC 2536.  

Figure 6 plots the [5.8]$-$[8.0] color against the [4.5]$-$[5.8] 
color.  Figure 7 compares the [4.5]$-$[5.8] color with
[3.6]$-$[4.5].  Also included in these two figures are the
predicted IRAC colors for interstellar dust \citep{li01}, the Sloan
Digitized Sky Survey quasars in the Spitzer Wide-Area Infrared
Extragalactic Survey (SWIRE) Elais N1 field \citep{hat05}, and the
colors of M0 III stars from M. Cohen (2005, private communication) and
field stars from \citet{whit04}.  The quasars have redshifts between
0.5 and 3.65; since their spectral energy distributions are power
laws, their infrared colors do not vary much with redshift.  From
these figures, it can be seen that clumps 23 and 26 have colors
consistent with those of quasars and field stars respectively and may
not be part of Arp 82.

From Figures 6 and 7, it can be seen that most of the clumps have
[3.6]$-$[4.5] colors similar to the colors of stars, (0.0$\pm$0.5
magnitudes) (M. Cohen 2005, private communication; \citealp{whit04})
implying these bands are mostly dominated by starlight.  The
[4.5]$-$[5.8] colors are generally very red, except for the two low
S/N clumps in the tail, 23 and 26.  Most of the clumps have
[4.5]$-$[5.8] colors between those of the ISM and stars, indicating
contributions from both to this color.  Most of the [5.8]$-$[8.0]
colors are between $\sim$1.6 and 2, similar to that of the ISM
\citep{li01}. Thus these bands are partially dominated by interstellar
matter.  Clump 24, which is in the northern tail, has colors similar
to those of ISM (upper limits are shown).  Thus, this appears to be a
very young star formation region with little underlying old stellar
population.

Figure 8 plots the ratio L(\ha)/L(3.6 \mic) against 8.0 \mic\
magnitude and Figure 9 compares the ratio L(\ha)/L(8.0 \mic) with 8.0
\mic\ magnitude.   Here and throughout we define L($\lambda$) =
$\Delta\lambda$L$_{\lambda}$($\lambda$) and $\Delta\lambda$ is taken
as the width of the band (IRAC Data Handbook).  The L(\ha)/L(3.6 \mic)
ratio is a proxy of the instantaneous star formation
normalized by an effective stellar  mass of the knots.  Note that
there is considerable scatter in both ratios, even in L(\ha)/L(8.0
\mic), in spite of the fact that both \ha\ and  8 \mic\ trace star
formation.  Note that clump 22 (at the end of the northern tidal
tail) and clump 21 (at the beginning  of the northern tidal tail)
have significantly higher \ha/mid-infrared ratios than the other
clumps.  Clumps 23 and 26 (in the northern tidal tail region) were
below our detection threshold in \ha.

To test whether the variations in L(\ha)/L(3.6 \mic) and especially
L(\ha)/ L(8.0 \mic) are due in part to extinction of the \ha\
emission, we plot the ratios against L(IR)/L(FUV) (Figures 10 and 11),
where L(IR) is the total infrared luminosity.  L(IR)/L(FUV) is a good
proxy for  extinction (see \S 4.2).   The L(IR), of the clumps was
estimated from the measured 8.0 \mic\ and 24 \mic\ fluxes using the
relation log[L(IR)] = log[$\nu$L(24)] + 0.908 + 0.793
log[L$_{\lambda}$(8.0) / L$_{\lambda}$(24)] \citep{cal05}.  Note that
$\nu$L(24) is the monochromatic luminosity with frequency
$\nu=1.27\times10^{13}$ Hz.  There is a scatter of $\pm40\%$ in the
L(IR) relation \citep{cal05}.  Weak anti-correlations may be present,
especially in the L(\ha)/L(8.0 \mic) plot, in that the regions with
lower dust obscuration tend to have higher L(\ha)/L(8.0 \mic) ratios.
There is, however, a lot of scatter in these plots, suggesting that
some of the variations in both the L(\ha)/L(3.6 \mic) and the
L(\ha)/L(8.0 \mic) ratios are intrinsic.  This scatter may be due to
PAH excitation by  non-ionizing photons contributing to the 8.0 \mic\
emission.

\subsection{Reddening and Ages}

The long wavelength IR light is due to the re-processing of UV light,
therefore L(IR)/L(FUV) is a measure of FUV opacity or dust optical
depth.  \citet{cal00} found that the UV obscuration at 1600 \AA\ is
related to the UV opacity by A(0.16\mic)=2.5 log[1/E F(IR)/F(FUV) +1],
where E=0.9 is the ratio of the bolometric correction of the
UV-to-near-IR stellar light relative to the UV emission at 1600 \AA\
and the dust bolometric correction to the fraction of FIR light
detected by the IRAS window.  The UV obscuration is related to the
extinction by A(0.16\mic)=4.39 \ebv\ \citep{cal00}.  Given the above
assumptions, the \ebv\ of the clumps in Arp 82 range from
$\sim$0.3$\pm0.1$ mag to $\sim$1.2$\pm0.1$ mag, with a mean of
$\sim$0.6$\pm0.1$.  The uncertainties in these \ebv\ estimates reflect
only the scatter in the L(IR) calibration.  Figure 12 is a plot of the
log[L(IR)/L(FUV)] versus distance from the NGC 2536 nucleus.  It can
be seen from Figure 12 that the reddening is generally greater  in NGC
2536 and the spiral region, while it is lower in the  bridge and tail
regions.

We have generated a set of model cluster spectral energy distributions
(SED) from the stellar population synthesis model Starburst99 (SB99)
code version 5.0 \citep{lei99}, and convolved these with the GALEX and
SARA bandpasses.  The SB99 models were generated assuming
stellar+nebular continuum emission, instantaneous starbursts,  solar
metallicity, a Salpeter IMF from 0.1 to 100 \mass, and a total mass of
\smass.  Changes in the evolutionary tracks due to metallicity are
small compared to our observational uncertainties, so other abundances
were not considered.  A second set of models were generated assuming
continuous star formation at a rate of 1 \massyr.   All other
assumptions in this second model are the same as the first.

The IRAC images suggest that a significant amount of dust is
associated with the clumps.  To correct the models for dust
obscuration, we reddened  them from 0-1.2 mag in 0.1 mag increments
assuming the \citet{cal94} starburst reddening law.  The assumption of
instantaneous or continuous star formation is quite important.  Ages
determined assuming continuous star formation may be factors of 10
larger than those assuming instantaneous star formation.  The bright
FUV, 8 \mic, and 24 \mic\ clump emission suggest a relatively young
stellar population, so instantaneous star formation is our working
assumption for the clumps.  Once the model IMF, abundance, and star
formation type is defined, the location of the clumps on a color-color
diagram depend only on the age and reddening.

Figure 13 is a histogram of the measured FUV/NUV flux ratios of the
clumps.  The median uncertainty in the ratios is given.  At the top of
Figure 13 are cluster ages determined from the SB99 code assuming no
reddening (blue diamonds), \ebv$=0.2$ mag (green triangles), and
\ebv$=0.6$ mag (red squares).  The age and reddening degeneracy can be
seen in this figure.  To help break the age/reddening degeneracy we
plot a color-color diagram.  Figure 14 is a plot of the log(FUV/R)
against log(FUV/NUV) with an SB99 model reddened with
\ebv$=0.0$(blue), 0.2(green) and 0.6(red) mag according to the
\citet{cal94} reddening law.   It can be seen from this figure that
the clumps fit nicely between the  \ebv$=0.2$ and \ebv$=0.6$ models.

To determine the ages and \ebv s of the individual clumps, we defined
the broadband age as the age associated with the log(FUV/NUV) flux and
the log(FUV/R) flux of a model point that was closest to the actual
data point for each clump in Figure 14.  The broadband \ebv\ is the
reddening applied to the above model.  Table 4 lists the broadband
ages and extinctions of clumps.  We also determined the ages of the
clumps from the H$\alpha$ equivalent width (\ewha) and SB99 (Table 4).
The \ewha\ traces the ionizing radiation from massive young O stars.
These ages will be referred to as \ewha\ ages.  The \ewha\ is in
principle a robust probe of clump age, as it is not sensitive to the
effects of reddening.  However, the \ewha\ is sensitive to the
determination of the continuum and our assumption about the
[N~{\scriptsize II}] emission.  The uncertainties in the ages in table
4 reflect only the uncertainties in the measured fluxes.  The mean
uncertainties on the broadband ages are $+34$ Myr and $-9$ Myr, $\pm1$
Myr on the \ewha\ ages, while the mean uncertainty on \ebv\ is
$\pm$0.1 mag.

With the exception of the two oldest clumps in the galactic nuclei
(clumps 2 and 16), the  broadband ages for the Arp 82 clusters range
from $\sim$1 to $\sim$30 Myr.  Clumps 2 and 16 are considerably older.
The cluster ages determined from the \ewha\ range from $\sim$6 to
$\sim$17 Myr and agree well with the broadband ages.  The FUV/NUV and
FUV/R trace stars from type B and A and type G and K, respectively;
thus they are a better tracer of slightly older clusters than
H$\alpha$.   The mean broadband \ebv=0.4 is similar to the mean \ebv\
determined from the L(IR) as described previously.  From the broadband
\ebv\ determinations it can be seen that the extinction is generally
greatest in the bridge region and the spiral region, consistent with
the extinction determined with the far infrared estimate.

Figure 15 plots the ages of the clumps against distance from NGC 2536.
The symbols are the same as before, with the blue symbols being the
broadband ages and the red symbols are the \ewha ages.  
The mean broadband ages of NGC 2536, the bridge region, the
spiral region, and the tail region are 30 Myr, 5 Myr, 15 Myr and 8 Myr
respectively.  The mean \ewha\ ages of NGC 2536, the bridge region,
the spiral region, and the tail region are 9 Myr, 11 Myr, 9 Myr, and
13 Myr, respectively.  Clumps 23 and 26, which may not be part of
Arp 82 (see \S 4.1), were below our detection threshold in \ha.  From
the  broadband age analysis it appears that the youngest clumps are
generally, but not exclusively,  in the bridge and tail regions.
Clumps 12, 14, and 15 make up the bright ``ocular ring'' at the outer
edges of the spiral region in NGC 2535.  These clumps are among the
youngest in the spiral region with broadband ages 17, 2, and 9 Myr
respectively, and \ewha\ ages of 7 Myr.

\subsection{Dust and the SFR}
 
The IR luminosity is the re-emission of absorbed UV photons, so the
sum of the UV and IR luminosities is  a good proxy for the total UV
emission and therefore is proportional to the star formation rate
(SFR).   In Figure 16, we plot the log[L(IR) + L(FUV)] against UV
opacity, log[L(IR) / L(FUV)].  A correlation can be seen.  In general,
as the dust opacity increases so does the log[L(IR) + L(FUV)].  This
correlation indicates that the dust opacity, and therefore the amount 
of interstellar matter, plays an important role in
the SFR.   Such a correlation has been seen with several other
galaxies \citep{hec98}.  \citet{cal05} find a correlation at the
$5\sigma$ level for the clumps in M51.  Figure 17 is a plot of the
log[L(IR)/L(FUV)] vs log[FUV/NUV].  It has been shown for starburst
galaxies that the UV dust opacity correlates with the UV colors
\citep{meu99}.  For Arp 82 there is a weak correlation at best in the
log[L(IR)/L(FUV)] vs log[FUV/NUV] plot.  Several of the clumps in Arp
82 are forming stars in the environments of tidal structures, where
the physical processes may be different, and likely more diverse, than
in nuclear starbursts.

We have determined the total SFR of the clumps using three
calibrations.  First, we used the relation, SFR(\massyr)$= 4.5 \times
10^{-44}$L(IR)(erg s$^{-1}$) \citep{ken98}, using L(IR) bootstrapped
from the 8.0 \mic\ and 24 \mic\ luminosities (\S 4.1).  The clumps
have a total SFR$_{IR}$ of $\sim$2.0$\pm0.8$ \massyr.   The
uncertainty reflects the scatter in the L(IR) calibration.  Second,
for comparison,  we used the relation SFR(\massyr)$= 1.4 \times
10^{-28}$L$_{\nu}$(FUV)(erg s$^{-1}$ Hz$^{-1}$) \citep{ken98}.  The
total SFR$_{UV}$ of the clumps is $\sim$0.6$\pm0.1$ \massyr,  slightly
lower than the SFR$_{IR}$.   Third, also  for comparison, we used the
relation SFR(\massyr)$=7.9 \times10^{-42}$L(H$\alpha$)(erg s$^{-1}$)
\citep{ken98}.  The total SFR$_{H\alpha}$ of the clumps is
$\sim$1.3$\pm0.2$ \massyr, slightly larger than the UV determination
and consistent with the SFR$_{IR}$.  The uncertainties in the
SFR$_{UV}$ and SFR$_{H\alpha}$ reflect only the uncertainties in the
measured fluxes.  No reddening correction has been applied to the
L$_{\nu}$(FUV) or L(H$\alpha$).    The SFR$_{IR}$ is probably the most
reliable because it is not nearly as sensitive to extinction effects.
However, the IR may include dust heating by non OB stars, even in the
clumps, and may overestimate the SFR.  \citet{cal05} find that L(IR)
is not directly proportional to extinction corrected Paschen-$\alpha$
in the M51 clumps.  The SFR$_{IR}$ is also sensitive to  the L(IR)
calibration discussed above.  The SFR$_{IR}$ of the entire Arp 82
system is $4.9\pm2.0$ \massyr,  vs $2.4\pm0.4$ for SFR$_{H\alpha}$.
The total clump SFR$_{IR}$  accounts for about 40\% of the entire
system SFR$_{IR}$.

Figure 18 plots the log[L(IR)+L(FUV)] of the clumps against the distance
from the NGC 2536 nucleus.  From this figure it can be seen
that  the SFR is greatest in the spiral region of NGC 2535 and in NGC
2536,  with much less star formation in the bridge and  tail regions.
The lower SFR of the clumps away from  the central regions could be
due to the fact that less gas was dragged there as a result of the
interaction.  Less gas spread over a still large volume implies lower
densities, lower  compressions, and larger clump masses to pull
together gravitationally.  Also, more turbulence in the tidal
structures may require larger masses for the self gravity to overcome
the turbulent pressure.  

\subsection{Masses of the Clumps}

From the measured R band fluxes and the ages and extinctions implied
by the broadband colors, we have determined masses for each of the
clumps in Arp 82 using our SB99 models.  These masses of the clumps
are given in Table 5.  The first column is the clump ID. The second
column is the region the  clumps can be found in and the third column
is the mass determined by this method (mass$_{R}$).  For comparison,
in Table 5 we also provide the  masses of the clumps using the FUV
flux (column 4).  The masses include stars from  0.1 to 100 \mass.
The uncertainties in the clump masses reflect only the uncertainties
in the measured fluxes.  It can be seen from the large uncertainties
that the FUV band is not a reliable mass tracer, as one would expect.
The R band is a more reliable tracer of mass than the FUV band because
the more populous lower mass stars contribute more to the R band flux.

The more massive clumps tend to be found in the spiral region while
the least massive clumps are in the tidal features.   The clumps have
a median mass$_{R}$ of about $8\times10^{7}$ \mass.  The ten tidal
clumps make up only about 3\% of the total clump mass$_{R}$.  The 2
clumps in the small companion, NGC 2536, and the 2 largest clumps in
the nucleus of NGC 2535 (13 and 16) make up about 82\% of the clump
mass$_{R}$.

As a test of the validity of our mass$_{R}$ determinations, we plot
the 3.6 \mic\ luminosity against mass$_{R}$ (Figure 19).  A strong
correlation is seen in this figure.  In Figure 14, it was shown that
the  3.6 \mic\ band is dominated by stars, so clump fluxes in this
band are a good proxy for mass.
  
We determined the mean mass to light ratios of the clumps,
$\alpha$=(M/M$_{\odot}$)/(L/L$_{\odot}$), using the mass$_{R}$  and
the R band luminosities.  We determined L/L$_{\odot}$ by assuming a
solar absolute R magnitude of 4.46 \citep{bel01}.  We find that the
mean clump mass to light ratio is $\alpha=0.77$ and the median
$\alpha=0.61$.  The clump mass to light ratios are plotted against
distance from NGC 2536 in Figure 20.

\subsection{H~{\scriptsize I} Arc}

An interesting tidal feature is clearly visible in the southeast of
the FUV image (Figure 2).  This `arc' is also visible in the
H~{\scriptsize I} maps of \citet{kau97} and was not detected in their
optical B and I images.  \citet{kau97} concluded that this feature was
a gaseous structure and suggest that it is a wake in the gas produced
by the passage of the companion  within or close to the extended
H~{\scriptsize I} envelope.  Its clear presence in the UV images shows
it does have a stellar component with fairly young stars.   Other
UV-bright tidal features that are optically faint and coincident with
H~{\scriptsize I} density enhancements have been observed with GALEX
(see for example,  \citealp{nef05}).  A previously undetected tidal
feature in NGC 4435/8 was discovered by \citet{bos05} with GALEX.

In addition to the 26 clumps discussed above, we identified 4 clumps
in this H~{\scriptsize I} arc.  These clumps are numbered 27, 28, 29
and 30 (Figure 2).  The clumps were identified visually on the FUV
image.  We imposed the same detection criterion as described above.
All 4 clumps were detected in the FUV and NUV.  Clumps 27, 28, and 30
were detected in H$\alpha$, clumps 27 and 30 were detected in R band
and 8.0 \mic, while only 30 was detected in 24 \mic.  None of the
H~{\scriptsize I} arc clumps were detected in 3.6 \mic, 4.5 \mic, or
5.8 \mic.

The H~{\scriptsize I} column densities for the clumps in this arc
range from $6\times10^{20} - 1\times10^{21}$ cm$^{-2}$ \citep{kau97}.
By comparing the FUV/NUV and FUV/R colors of the H~{\scriptsize I} arc
clumps 27 and 30 to SB99 we find that they have broadband ages of
$7^{+83}_{-0}$ and $6^{+74}_{-3}$ Myr respectively and
\ebv$=0.1^{+0.0}_{-0.1}$ and $0.2^{+0.1}_{-0.2}$ mag respectively.
This very low extinction is consistent with the apparent lack of dust
suggested by the low IR fluxes.  Clumps 27 and 30 have a mass$_{R}$ of
$0.32^{+4.24}_{-0.10}\times10^6$ \mass\ and
$0.52^{+6.95}_{-0.30}\times10^6$ \mass\ respectively.

\subsection{The Underlying Stellar Population}

In sections 3 and 4, we have extensively discussed the young
star-forming clumps in Arp 82.  To investigate the underlying stellar
population, we subtracted the total clump flux from the total flux in
each waveband.  We  estimated the age and extinction of this
underlying population by comparison to a set of SB99 models similar to
that described in \S 4.2.  Because an older population could not be
effectively modeled with the assumption of an instantaneous burst, we
instead generated a model assuming a continuous SFR of 1 \massyr.  All
other model assumptions were the same as before.  From the FUV/NUV and
FUV/R ratios  and SB99 we found that the underlying population was
best matched to a model with \ebv$=0.3$ mag and an age of $\sim2$ Gyr.   
The \ebv\ estimated here is similar to the extinction we determined 
for the clumps.

This suggests that the underlying diffuse stellar component in Arp 82
is  dominated by a $\sim2$ Gyr population and that the progenitors may
have been of very late type or low surface brightness galaxies with
very modest underlying old stellar populations.  It should be noted
that the FUV/NUV and FUV/R ratios are more sensitive to young and
intermediate age stars than truly old stars.  It is possible there is
an older underlying population in Arp 82 that we are not able to
detect.  Without the aid of near-IR bands that reflect the peak
contributions from M stars, we cannot rule out the presence of a truly
old, $\sim$10 Gyr, population.  Future work with proposed near-IR maps
and optical spectra will further  constrain the age of the underlying
population.

\section{COMPARISON WITH A NUMERICAL SPH MODEL OF THE INTERACTION}

The distinctive morphology of this system provides strong constraints
on any numerical model of the interaction.  Arp 82 is an M51-type
system with a strong bridge as observed in the optical/IR bands and in
the 21 cm radio band, and a long countertail stretching to the
northwest. These features, together with the ocular waveform
(\citealp{elm93,kau97}) in the primary disk suggest that the encounter
is strongly prograde for the primary galaxy. That is, the companion
orbits the primary in the same sense as the primary rotates, and the
companion orbital plane is close to the primary disk plane. These
features also constrain interaction time scales. On one hand, an
ocular wave is a relatively short-lived feature, though it also takes
some time to develop (see e.g., the models of the NGC 2207/IC 2163
system of \citealp{str05}).  On the other hand, the length of the
primary tail suggests that this interaction has been underway for some
time.  This is affirmed by the H~{\scriptsize I} observations
\citep{kau97}, which indicate that not only has there been time to
form the bridge, but also time to transfer significant amounts of gas
through the bridge to the companion. Finally, the mix of young and
intermediate age populations yielded by the population synthesis
modeling suggests that the interaction has been underway for a time
comparable to the age of the intermediate age populations.

At the same time, the observed FUV/NUV and FUV/R ratios suggest that
the star formation in the tail and the bridge, is recent. The clumps
in the tail are among the youngest with ages of less than 70 Myr, and
perhaps as young as a few Myr. This is much less than the likely
kinematic time scale for the arm to reach its current distance from
the center.  The companion is very disturbed, both morphologically and
kinematically, so we cannot derive much information about the
interaction from its structure. However, we would expect a successful
model to reproduce its disturbed structure, and intense star formation
activity.

The H~{\scriptsize I} mapping of \citet{kau97} provides a good deal of
kinematic information on this system. However, for present purposes we
seek an approximate model that can help understand the range of
time scales indicated by large-scale morphological structures and the
general star formation history. We have not attempted to produce a
model that is sufficiently detailed to match the kinematics as well.

\subsection{Model Details}

The model described below was produced with the SPH code of Struck
(1997, also see Struck et al. 2005). This code employs some
simplifications to allow it to efficiently explore a range of
collision parameters. The foremost of these is that rigid dark halo
potentials are used, and hydrodynamic forces are computed on a grid
with fixed spacing. Local gravitational forces are computed between
particles in adjacent cells, to capture local gravitational
instabilities, which are important for modeling star formation. In
addition to the halos the model galaxies have disks consisting of gas
particles and collisionless star particles of equal mass.

In the models described below the following particle numbers were
used: 42900 primary disk gas particles, 5640 primary star particles,
13590 companion disk gas particles, and 2490 companion star
particles. The extensive gas disks, and limited stellar disks, used in
these models represent our belief that the progenitor galaxies were
likely of very late type or were low surface brightness galaxies, with
very modest old stellar populations.

We adopt the following scaling constants for the model: time unit $=
400$ Myr, length unit $= 2.0$ kpc, and mass unit $M_h = 3.9\times10^{10}
M_{\odot}$. With these scalings we obtain a peak rotation velocity of
about 200 km s$^{-1}$ in the primary, which reasonably matches the
uncertain observational constraints \citep{kau97}.

The mass ratio of the two galaxies can be estimated by comparing the
mass of each within a radius of 20 kpc; it is 0.13 \citep{kau97}. The
form of the rigid potential used gives a test particle acceleration of,

$$a = \frac{G{M_h}}{{\epsilon}^2}\ \frac{r/{\epsilon}}{(1 +
r^2/{\epsilon}^2)^{n_h}},$$

\noindent
where $M_h$ is a halo mass scale, $\epsilon$ is a core radius (set to
2.0 and 4.0 kpc for the primary and companion, respectively), and the
index $n_h$ specifies the compactness of the halo. For the primary we
use $n_h = 1.1$, which gives a slightly declining rotation curve at
large radii. For the companion we take $n_h = 1.35$, which gives a
declining rotation curve at large radii. We have not undertaken a
detailed examination of the effects of different values for these
exponents. Because the two galaxies have different halo potentials,
the effective mass ratio (i.e., the ratio of masses contained within
radii equal to the galaxy separation) is large for small separations,
and small for large separations. This effect makes encounters between
the two galaxies quite impulsive. The model presented here includes
the effects of dynamical friction with a Chandrasekhar-like frictional
term \citep{str03}. The effects of this term are small except
for brief times near closest approach. It ultimately draws the two
galaxies into a merger at times shortly past the present (the second
close encounter).

The initial disk sizes of the primary gas disk was 19.2 kpc, and the
companion gas disk was 14.4 kpc.  These values, and the initial
orientation of the companion, are fairly arbitrary; we have not
attempted to optimally fit them. The primary disk was initialized in
the computational x-y plane. The companion disk was initialized in the
x-y plane and then tilted $20^{\circ}$ around the y-axis such that
positive x values had positive z values. It was then tilted
$-25^{\circ}$ in the x-y plane around the y-axis. After the run the
whole model x-y was rotated by $-90^{\circ}$ in the x-y plane around
the primary center to better fit the observed orientation. The initial
position and velocity vectors of the companion relative to the primary
center were: (0.0 kpc, 30.0 kpc, 4.0 kpc) and (200 km s$^{-1}$, -60 km
s$^{-1}$, -25 km s$^{-1}$).

\subsection{Model Results}

Figure 21 shows various views of the model results at a time near the
present, as judged by the morphology. All axis labels are in kpc with
the adopted model scaling. The top two panels show the relative orbit
of the companion as a dashed curve from the beginning of the run and
into the future, with the cross marking the initial position. The top
two panels also show the distribution of cold (or unheated) gas
particles from the two disks at the present time. The bottom right
panel shows gas particles in which the threshold density has been
exceeded, so star formation and heating have been turned on in
them. The lower left panel shows the distribution of hot gas
particles, i.e., particles in which star formation heating was on
recently, but which have not had time to cool. The initial position
and about the first three fourths of the trajectory are not strongly
constrained by observation, and are fairly uncertain.

The two initial gas disks were very similar, so the contrast at the
present time is quite striking. The most prominent tidal features, the
bridge and tail of the primary, were largely formed as a result of the
most recent close encounter. This encounter occurred at a position
angle of about $45^{\circ}$ in Fig. 21. The bold curve in Figure 22
shows the separation distance.  At the present time
the two galaxies have moved somewhat apart after close approach. The
upper left panel of Figure 21 suggests that after this time the two
will merge.

Figure 21 shows that this model does a pretty good job of
reproducing the extensive gas bridge observed in this system (also see
the earlier model of \citealp{kla93} and \citealp{kau97}), and the long,
fairly narrow tidal tail. The optical tail is unusually long, about
three times the size of the primary disk. The model tail is only about
two times the size of the star-forming part of the primary disk, which
may indicate that the initial model gas disk should have been larger.

There was an earlier close encounter near the beginning of the
run. The primary also formed a bridge and tail after this
encounter. However, these features were weaker, and did not persist or
leave much sign of their existence in the current morphology.

This is not true in the case of the companion. The initial encounter
was close enough and the orientation of the companion was such that a
strong radial (ring-like) wave was generated. This, and subsequent
tidal forces, resulted in much of the outer gas disk of the companion
being pulled off in a long and very diffuse plume. Initially, the gas
retained by the companion expanded outward in the ring wave, but most
of it fell back in close to the time of the most recent close
encounter. The two effects combined to put a large mass of compressed
gas in the core of the companion and trigger repetitive starbursts
(Fig. 22).

We have attempted to reproduce a plausible star formation history of
both galaxies with the models.  The results are shown in Figure
22. First of all, the disk structure, gas particle mass (about
$1.3\times10^5 M_{\odot}$), and the star formation threshold density
were all initialized such that the two gas disks would have a low rate
of star formation in isolation. The dotted curves in Fig. 22 show the
star formation history of the two disks run in isolation for a time of
slightly more than 2 Gyr. After some initial transients both disks
settle to a very low rate of stochastic star formation. In the
interaction simulation, the companion retains a low rate of star
formation up to the time of the most recent close encounter. As a
result of the tidal distortion and the repetitive starbursts the
companion could be converted from a very regular
low-surface-brightness disk to the compact, but irregular form
observed now.

The interaction is prograde for the primary galaxy, so even the weak
initial encounter triggers an enhanced rate of star formation in the
core of the primary, and somewhat later in induced spiral
waves. Nonetheless, even this enhanced rate is modest until the most
recent encounter, when a burst is triggered with about 10 times the
SFR of the isolated primary. The model further predicts that with
merger in the near future the SFR will rise precipitously. This
prediction cannot be checked, of course, but it is in accord with what
we know about gas-rich, infrared luminous, merging systems.

Nonetheless, this range of SFR amplitudes in the model is probably not
enough to account for the observation that the bulk of the star
formation (SF) has occurred in the last $\sim2$ Gyr.  Given the long
time before the interaction, even at the low quiescent rates shown in
Fig. 22, only of order 5\% of the stars would be produced in the
interaction with the rates shown there. It is likely that the true
quiescent SFRs were even lower, and the core burst rates in the first
interaction were much higher than in the model of Fig. 22.  The former
could be easily achieved by modifying the threshold density parameter
for SF feedback. Modeling stronger bursts would probably require
higher particle resolution, a more massive companion (to increase the
perturbation on the primary) and more sophisticated thermal
physics. Clearly, this system poses interesting tests for
phenomenological feedback models, though we believe more detailed
modeling should await spectral observations that provide stronger
constraints on the stellar population ages.

We note for completeness that if the model were run backwards from
its rather arbitrary initial conditions, it would have spent at least
another Gyr in an outward loop. This assumes that the adopted halo
potentials extend that far out, and that the loop is like the one
illustrated in Figs. 21 and 22, but somewhat larger due to the modest
effects of dynamical friction. It is equally likely that the halo
potentials start to fall off more rapidly at large separations, and
that the companion took a much longer time to come in from a much
greater separation. Then the initial encounter modeled here would have
been the first.

Finally, it is interesting to use Fig. 22 to compare the SFR to the
galaxy separation over the course of the simulation. Generally, these
two quantities correlate, but not monotonically. This is illustrated
by the echo bursts that occur near the present time, and well after
the last close encounter.

\section{COMPARISON TO OTHER GALAXIES}

How do the star forming properties of the Arp 82 clumps compare to
other galaxies studied in detail by Spitzer?  The [4.5]$-$[5.8] and
[5.8]$-$[8.0] colors of the Arp 82 clumps are typically redder than
those in Arp 107, the other SB\&T galaxy we have studied in detail
\citep{smi05}.  This indicates younger stellar populations on average
in Arp 82; consistent with its more quiescent IRAS properties.  The
IRAC colors of the clumps in Arp 82 are generally similar to those in
IC 2163 and NGC 2207 \citep{elm06}.  The log[FUV/NUV] distribution of
the clumps in Arp 82 resembles that of the clumps in M51
\citep{cal05}.  This suggests that the clump ages in the two systems
are likely similar.

Very young star forming regions have also been found in a myriad of
other  interacting galaxies using UV and optical studies.  For
example, \citet{wei04} find that 59\% of the star forming regions in
the advanced merger NGC 4194 are younger than 10 Myr, while
\citet{han06} report that 14 star forming regions in NGC 4194 are
$\sim$6 Myr.  \citet{han03} find that several knots in NGC 3395 and
NGC 3396 have ages less than 20 Myr; some are as young as 5 Myr.
\citet{egg04} reports that the ages of 12 knots in NGC 3994 are less
than 20 Myr and the ages of 22 knots in NGC 3995 are younger than 10
Myr.  In a study of NGC 1741, \citet{john99} found star forming
regions as young as a few Myr.   \citet{john00} report ages less than
10 Myr for nearly half the  knots in He 2-10.  The Antennae (NGC
4038/39) has clusters with ages ranging from $\sim$5-10 Myr
\citep{whi03}.  

Most of the star forming regions discussed in the above galaxies are
proto-globular cluster candidates.  Some of the clumps in Arp 82 are
similar in age to these star forming regions, but with the  exception
of a few clumps in the tail, are typically too massive to be
proto-globular clusters.  The mean mass$_{R}$ of the Arp 82 clumps is
$3.6\times10^8$ \mass, much more massive than typical Galactic
globular clusters, which have masses $\sim$10$^{5}-10^{6}$ \mass\
\citep{pry93}.    The mass of 15 clumps in IC 2163 and NGC 2207 were
found to range from $2\times10^4$ to $1\times10^6$ \mass\
\citep{elm06}.  \citet{cal05} find that the brightest H$\alpha$ knot
in M51 is $\sim$2$\times10^5$ \mass, while the UV emitting sources
range in mass from $\sim$6$-$8$\times10^7$ \mass.  The Tidal Dwarf
Galaxies (TDG) in the \citet{bra01} sample have masses ranging from
$2\times10^6$ \mass\ to $4.5\times10^8$ \mass, while the TDG
candidates in the \citet{hig05} sample have masses $\sim$2$\times10^7$
\mass\ to $\sim$3$\times10^8$ \mass, similar to the masses of most of
the clumps in  Arp 82.  However, clumps 22, 24, and 25 (in the Arp 82
tail) have masses of $2^{+22}_{-1} - 10^{+27}_{-5}\times10^6$ \mass\
which are similar to that of Galactic globular clusters and the UV
sources in M51.

The star forming clumps in Arp 82 are larger than globular clusters.
The mean full width at half maximum of the clumps in R band is $\sim$1
kpc.  This size is much larger than typical Galactic globular clusters
which have effective radii of about 3 pc \citep{mil97}.  The clumps
have sizes and masses more like giant star forming complexes  and TDGs
than globular clusters.  Given the limited resolution of these data,
it is possible that some of the clumps are comprised of multiple
unresolved clusters.  \citet{elm06} found that the Spitzer clumps in
the IC 2163 ocular, which  are similar to ours, were resolved into
smaller clusters by HST.

For a sample of Galactic globular clusters, \citet{pry93} find mass to 
visible light ratios that range from 0.5 to 6.2 with a mean ratio of
$\sim$2.4.  Although the mean mass to light ratio we derive for the
clumps is within this range, it is far below the \citet{pry93} mean.
Figure 20 is a plot of the mass to light ratios of the clumps against
distance from NGC 2536.

The SFR$_{IR}$ of the entire Arp 82 system is $4.9\pm2.0$ \massyr.
The SFR alone may not adequately suggest an enhancement in star
formation.  The [3.6]$-$[24] color is a  good measure of the ratio of
the SFR to the total stellar mass.  The total system [3.6]$-$[24]
color of Arp 82 is 5.9,  near the peak of the Arp disk distribution
and redder than the majority of the normal spirals in the Spitzer
infrared nearby galaxy sample (SINGS) (see \citealp{smi06}),
suggesting an interaction-induced  enhancement in star formation.

\section{SUMMARY}

We present a UV, optical, and mid-IR study of the star forming
properties of the interacting pair Arp 82, using data from GALEX,
SARA,  and Spitzer.  We have identified 30 star forming regions
(clumps).   The clumps contribute 33\% and 40\% of the total FUV and
NUV emission respectively and similar percentages (26\%, 26\%, 28\%,
33\% and 36\%) of the R band and IRAC bands (3.6, 4.5, 5.8, and 8.0
\mic\ respectively).   Larger percentages are found at 24 \mic\ and
H$\alpha$ (70\% and 55\%,  respectively).  The relatively low clump
contribution to the 8.0 \mic\ emission implies a considerable
contribution to the non-clump 8.0 \mic\ from PAH heating by
non-ionizing stars.

The clumps have [3.6]$-$[4.5] colors similar to the colors of stars.
Most of the  clumps have [4.5]$-$[5.8] colors between those of the ISM
and stars, indicating contributions from both to this color.  Most of
the [5.8]$-$[8.0] colors are similar to that of the  ISM. Thus these
bands are dominated by interstellar matter.  Clumps 23 and 26 have
colors consistent with those of quasars and field stars respectively
and may not be part of Arp 82.  There is considerable scatter in both
the L(\ha)/L(3.6 \mic) and L(\ha)/L(8.0 \mic) ratios of the clumps.
Some of the variation in the L(\ha)/L(8.0 \mic) ratios maybe due to
extinction, but some is intrinsic, perhaps due  to polycyclic aromatic
hydrocarbon (PAH) excitation by  non-ionizing photons.

The broadband ages of the clumps indicate that the majority are less
than 100 Myr and could be as young as a few Myr.  The clumps in
the tidal features tend to be younger.  The extinction across the
galaxy varies from \ebv $\sim$0.2$^{+0.1}_{-0.2}-0.7^{+0.2}_{-0.0}$
mag, being generally greater in NGC 2536 and the spiral region and
lower in the  bridge and  tail regions.  The masses of these clumps
range from a few $10^6$ to a few $10^9$ \mass, with the two nuclei
being by far the most massive, making up 82\% of the total clump mass.
The total SFR$_{IR}$ of Arp 82 is 4.9$\pm2.0$ \massyr, with $\sim$16\%
of the new stars forming in the two nuclei.

We have used an SPH code to model the interaction and reproduce the
star formation history.  The model indicates that the galaxies have
undergone two close encounters.  In the model, the primary suffered a
mild starburst during the first close encounter and a stronger burst
more recently, consistent with  the observations.  In the interaction
simulation, the companion retains a low rate of star formation up to
the time of the most recent close encounter.  The companion may have
converted from a very regular low-surface-brightness disk to the
compact irregular form observed now, as a result of the tidal
distortion and the repetitive starbursts.  The model predicts that
with merger in the near future the SFR will rise precipitously. This
prediction cannot be checked, of course, but it is in accord with what
we know about gas-rich, infrared luminous, merging systems.

The star formation history for this system derived from observations
and models, suggest that this is a very unusual system. The initial
formation of these two galaxies seems to have been as unspectacular as
the model star formation transients of the isolated disks shown in
Fig. 22. This quiescent initial formation was evidently followed by a
long period, only a little less the age of the universe, when very few
additional stars were produced. In this sense, galaxy formation got
stuck, before normal, late-type disks could be formed. The present
interaction and eventual merger may result in an early to intermediate
Hubble type disk galaxy.

Since the galaxies in this system are of intermediate mass, their
evolution is in rough accord with the downsizing concept, wherein
large galaxies are supposed to have formed at early times in dense
environments, and most of the current star formation is occurring in
much smaller galaxies (see e.g., \citealp{jun05,bun05, mat06} and
refs. therein).  Recent work further suggests that most of the stars
in intermediate mass galaxies formed 4-8 Gyr ago, and most current
star formation is occurring in still smaller galaxies (\citealp{ham05}
and refs. therein).  Arp 82 may be a downsizing outlier, and
could provide a local view of the ``age of intermediate mass galaxies." The
stellar populations in this relatively nearby system could be studied
in much greater detail than those of its higher redshift kin.

\acknowledgments

The authors thank the anonymous referee for their comments and suggestions.
This work is based in part on observations made with the Spitzer
Space Telescope, which is operated by the Jet Propulsion Laboratory,
California Institute of Technology under contract with NASA. 
GALEX is a NASA Small Explorer mission, developed in cooperation with the
Centre National d'Etudes Spatiales of France and the Korean Ministry of 
Science and Technology.  This research was supported by NASA Spitzer grant
1263924, NSF grant AST 00-97616, NASA LTSA grant NAG5-13079, and GALEX
grant GALEXGI04-0000-0026.  This work has made use of the
NASA/IPAC Extragalactic Database (NED), which is operated by the Jet
Propulsion Laboratory, California Institute of Technology, under
contract with NASA.

\newpage
\clearpage
\begin{figure*}[t]
\includegraphics[width=6.1in]{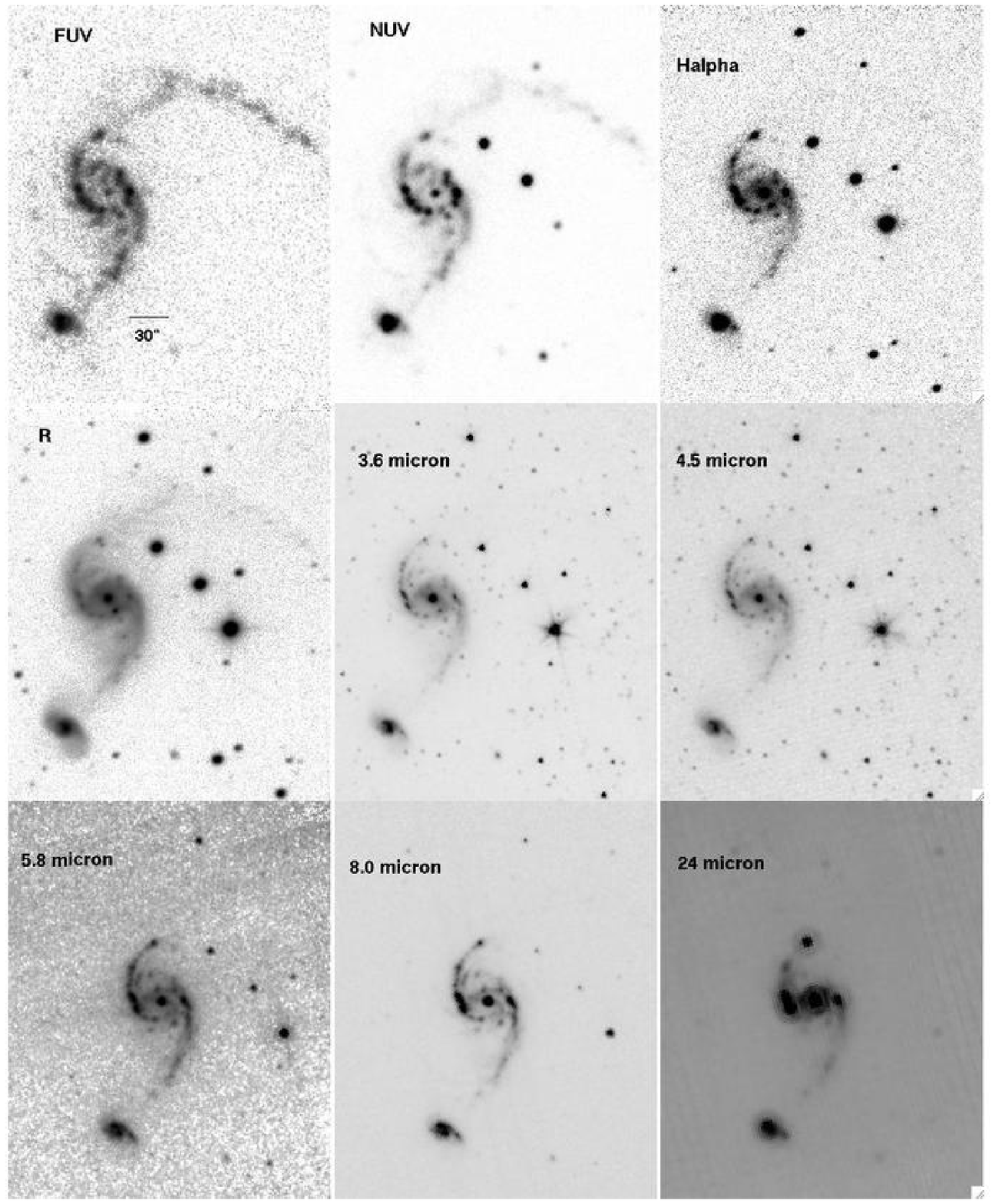}
\caption[f1.eps]{From top (left to right):  FUV, NUV, H$\alpha$, R, 3.6 \mic, 4.5 \mic, 5.8 \mic, 8.0 \mic, \& 24 \mic.  In all the images north is up and east is to the left.  The scale bar is 30\arcsec\ and corresponds to $\sim$8.2 kpc.  The large galaxy to the north is NGC 2535 and the smaller companion to the south is NGC 2536.  \label{f1}}
\end{figure*}

\newpage
\clearpage
\begin{figure*}[t]
\includegraphics[width=6.1in]{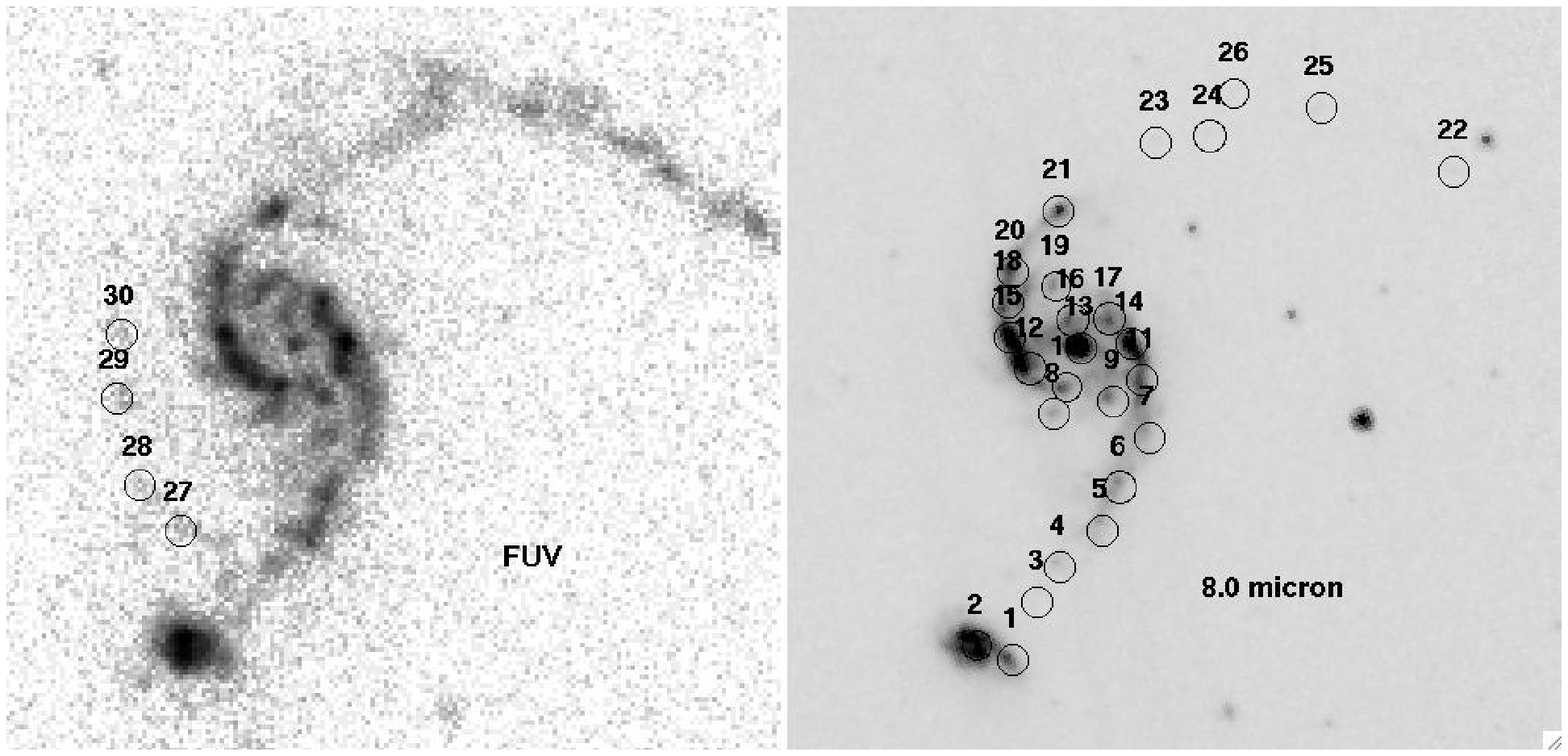}
\includegraphics[width=6.1in]{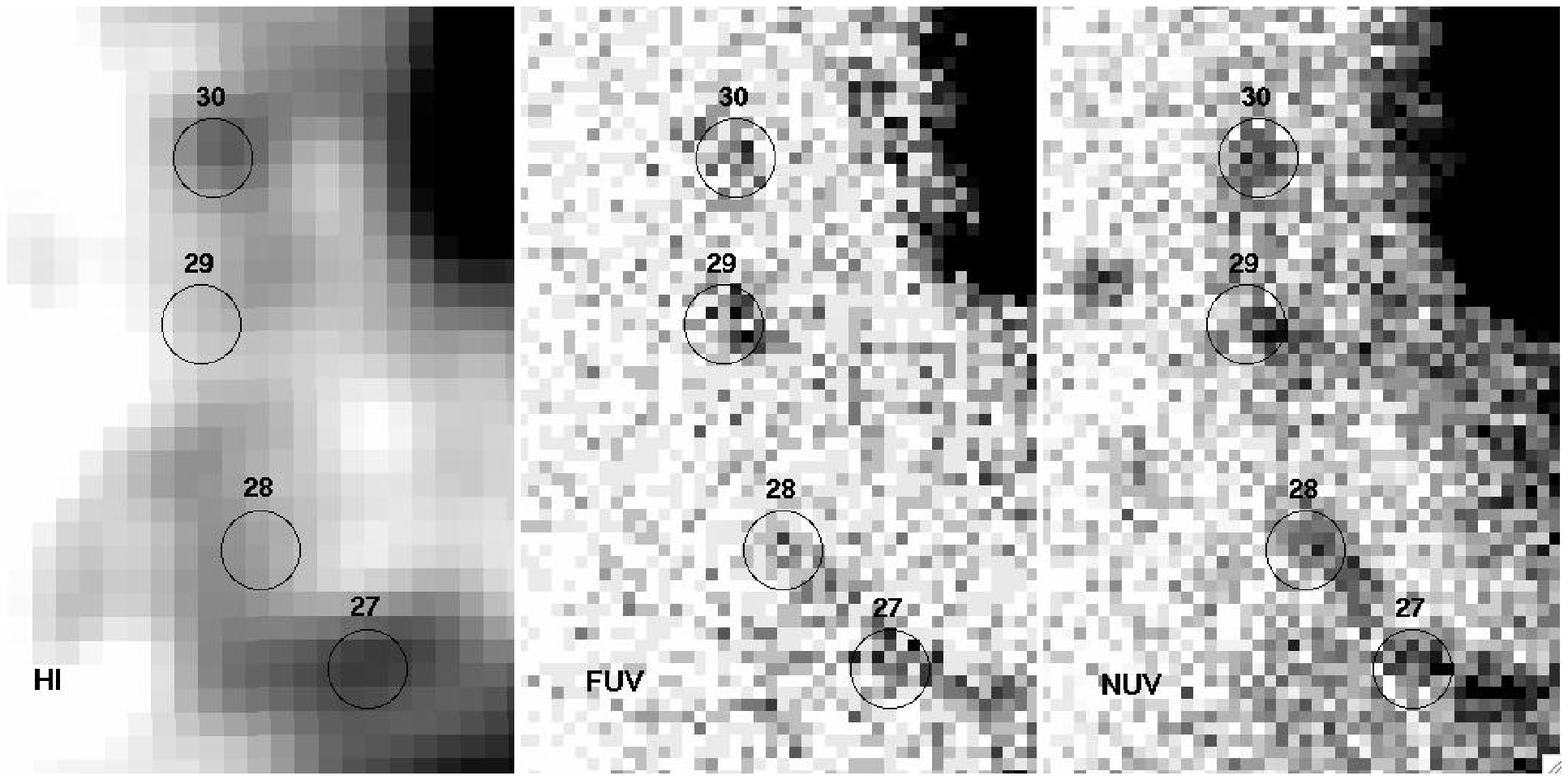}
\caption[f2a.eps,f2b.eps]{Expanded FUV (top left) and 8.0 \mic\ (top right) images with the clumps identified.   Clumps 27-30 lie along the H~{\scriptsize I} arc and are shown on the FUV image.  Zoomed in H~{\scriptsize I} map \citep{kau97}, FUV, and NUV images of the H~{\scriptsize I} arc (bottom left, middle, and right respectively). \label{f2}}
\end{figure*}

\newpage
\clearpage
\begin{figure}[ht]
\plotone{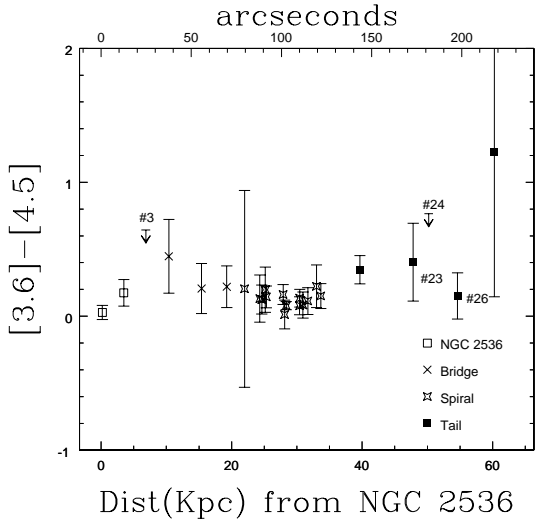}
\figcaption[f3.eps]{[3.6]$-$[4.5] gradient.  Open squares represent clumps in NGC 2536, x's represent clumps in the bridge, stars represent clumps in the spiral and filled squares represent clumps in the counter tail.  Clumps 3 and 24 are plotted as upper limits.  The mean [3.6]$-$[4.5] color for \citet{whit04} field stars is -0.05, while the predicted value for interstellar dust is -0.35 \citep{li01}.  \label{f3}}
\end{figure}

\begin{figure}[ht]
\plotone{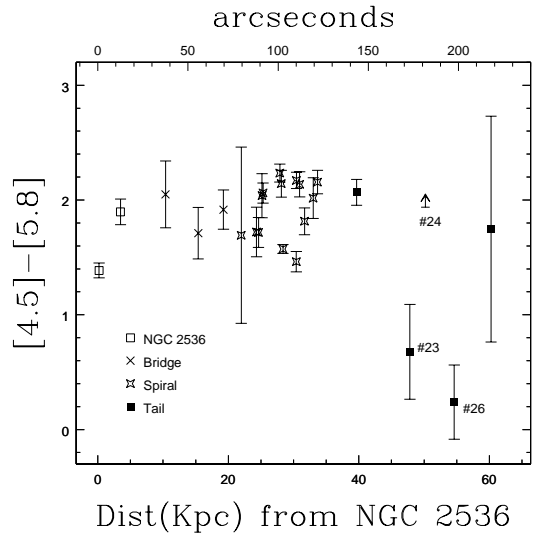}
\figcaption[f4.eps]{[4.5]$-$[5.8] gradient.  The symbols are the same as in Figure 3.  Clump 24 is plotted as a lower limit.  The mean [4.5]$-$[5.8] color for \citet{whit04} field stars is 0.1, while the predicted value for interstellar dust is 3.2 \citep{li01}.  \label{f4}}
\end{figure}

\newpage
\clearpage
\begin{figure}[ht]
\plotone{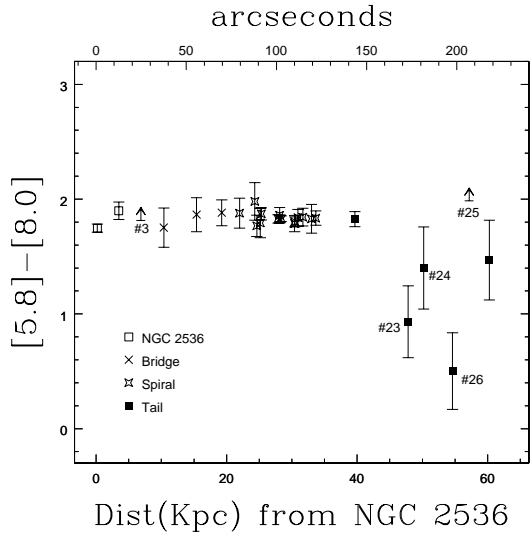}
\figcaption[f5.eps]{[5.8]$-$[8.0] gradient.  The symbols are the same as in Figure 3.  Clumps 3 and 25 are plotted as lower limits.  The mean [5.8]$-$[8.0] color for \citet{whit04} field stars is 0.05, while the predicted value for interstellar dust is 2.1 \citep{li01}.\label{f5}}
\end{figure}

\begin{figure}[ht]
\plotone{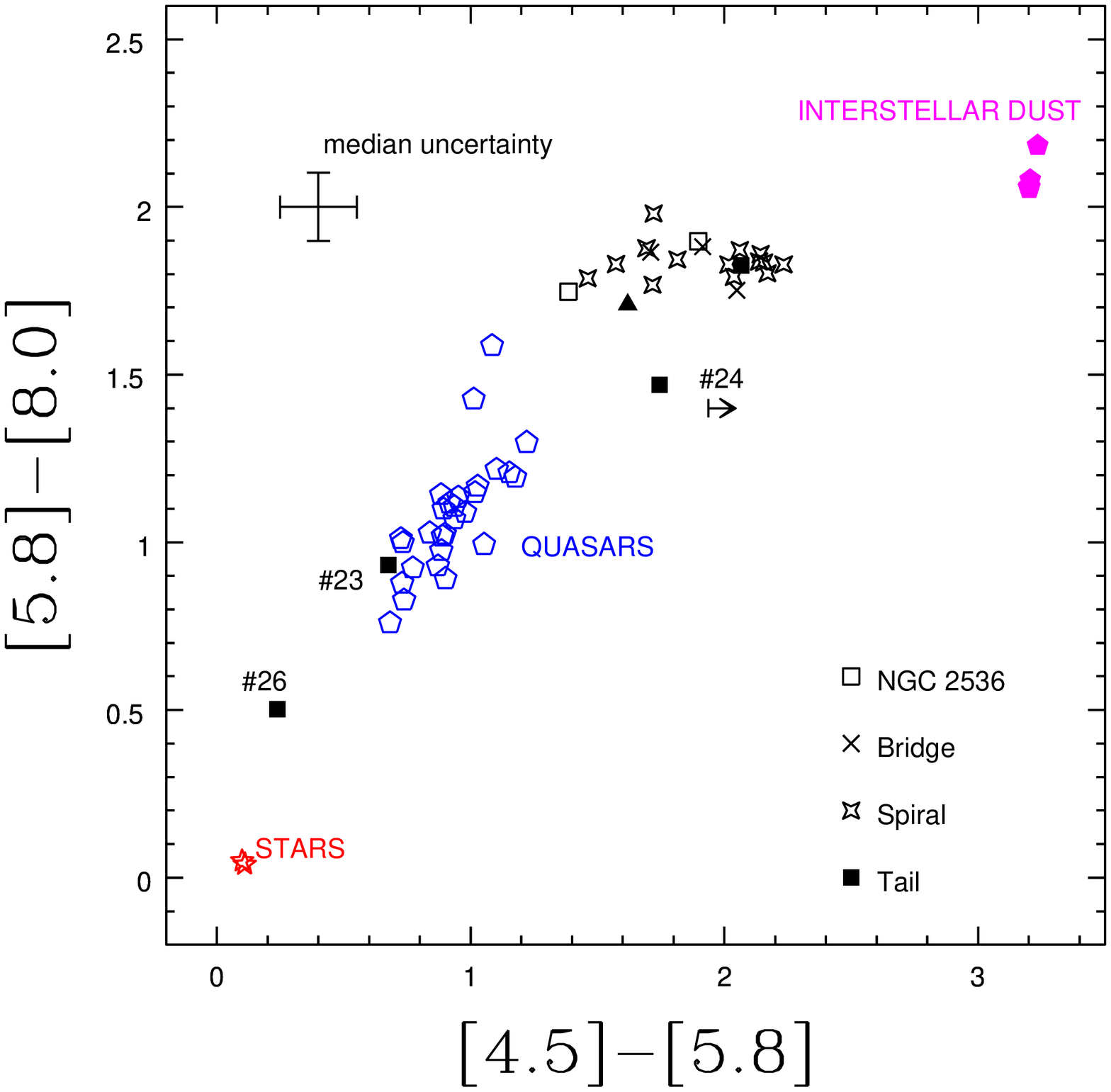}
\figcaption[f6.eps]{[5.8]$-$[8.0] vs [4.5]$-$[5.8].  The symbols are the same as in Figure 3.  The solid black triangle represents the entire Arp 82 system.  The solid pentagons represent predicted IRAC colors for interstellar dust \citep{li01}, open pentagons represent the \citet{hat05} colors of quasars, and the open 5 point stars represent the mean colors of M0 III stars from M. Cohen (2005, private communication) and field stars from \citet{whit04}. The median uncertainty is also shown.  Clump 24 is plotted as a lower limit.  \label{f6}}
\end{figure}

\newpage
\clearpage
\begin{figure}[ht]
\plotone{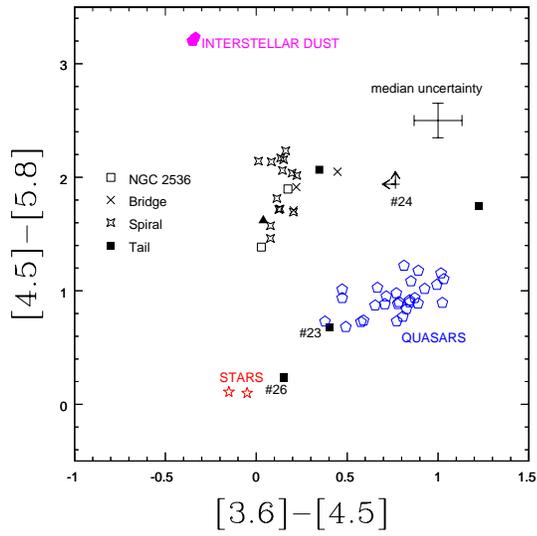}
\figcaption[f7.eps]{[4.5]-[5.8] vs [3.6]-[4.5].  The symbols are the same as figure 13.  Clump 24 is plotted as a lower limit.  \label{f7}}
\end{figure}

\begin{figure}[ht]
\plotone{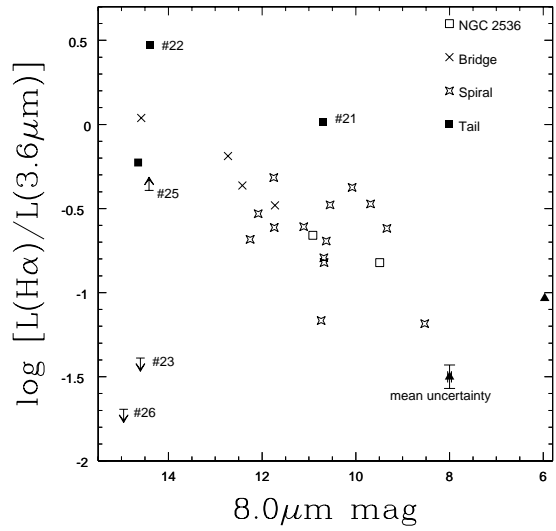}
\figcaption[f8.eps]{Log[L(\ha)/L(3.6\mic)] vs 8.0\mic\ mag.  The symbols are the same as in Figure 3.  The solid black triangle represents entire Arp 82 system.  \label{f8}}
\end{figure}

\newpage
\clearpage
\begin{figure}[ht]
\plotone{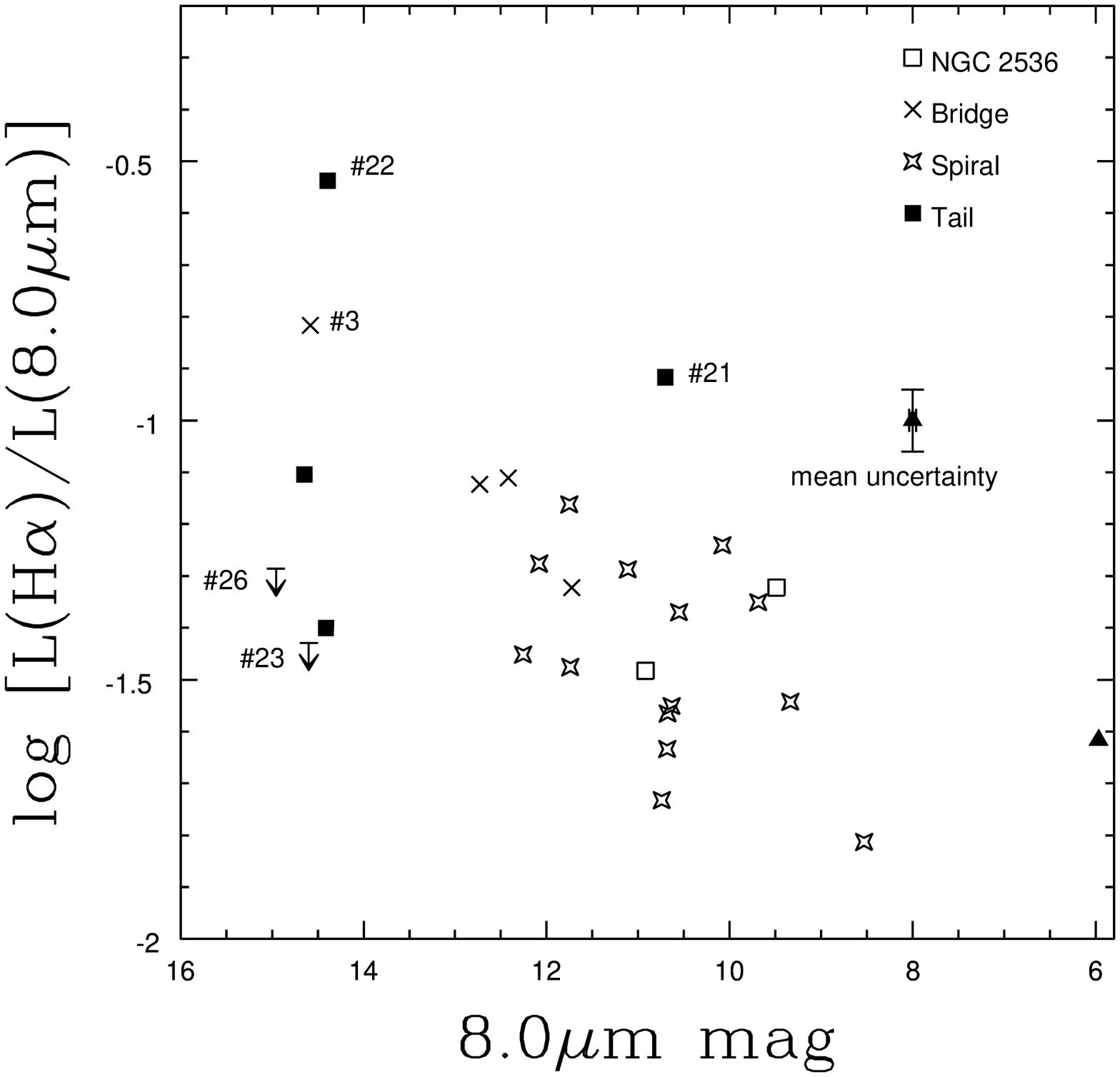}
\figcaption[f9.eps]{Log[L(\ha)/L(8.0\mic)] vs 8.0\mic\ mag.  The symbols are the same as in Figure 8.  \label{f9}}
\end{figure}

\begin{figure}[ht]
\plotone{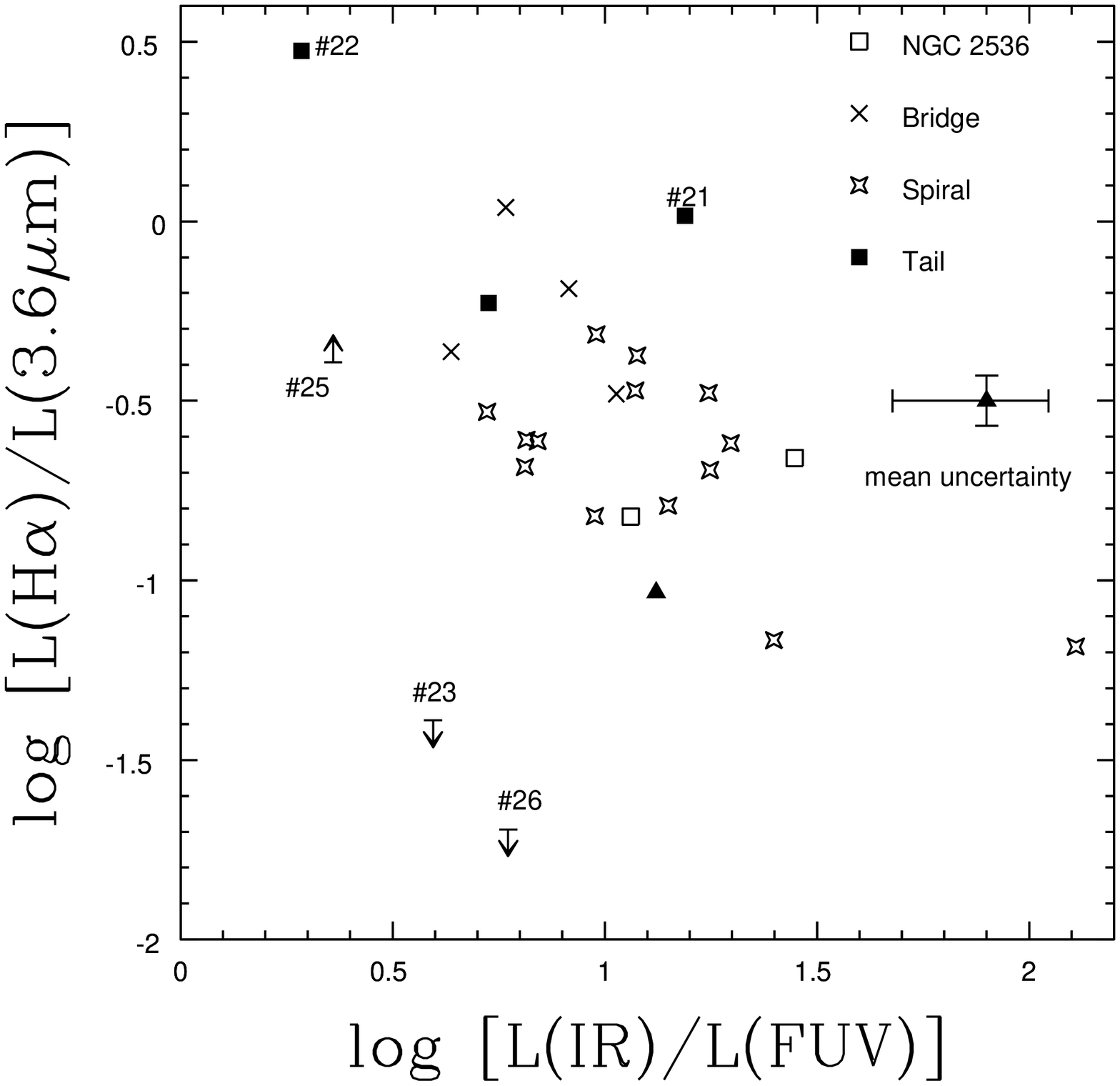}
\figcaption[f10.eps]{Log[L(\ha)/L(3.6\mic)] vs Log[L(IR)/L(FUV)].  The symbols are the same as in Figure 8.  The mean uncertainty for the horizontal axis reflects only the scatter in the L(IR) calibration.\label{f10}}
\end{figure}

\newpage
\clearpage
\begin{figure}[ht]
\plotone{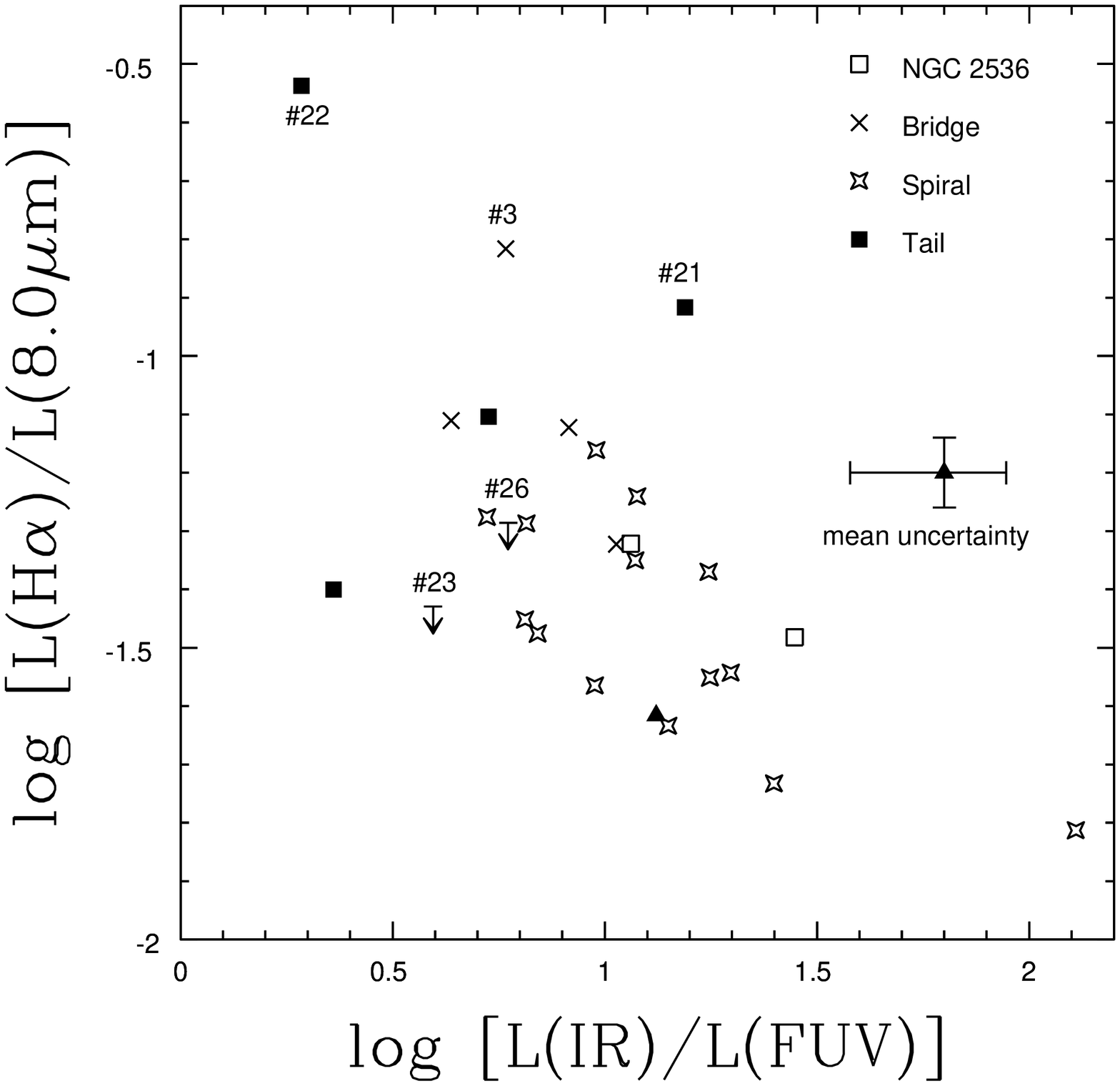}
\figcaption[f11.eps]{Log[L(\ha)/L(8.0\mic)] vs Log[L(IR)/L(FUV)].  The symbols are the same as in Figure 8.  The mean uncertainty for the horizontal axis reflects only the scatter in the L(IR) calibration.\label{f11}}
\end{figure}

\begin{figure}[ht]
\plotone{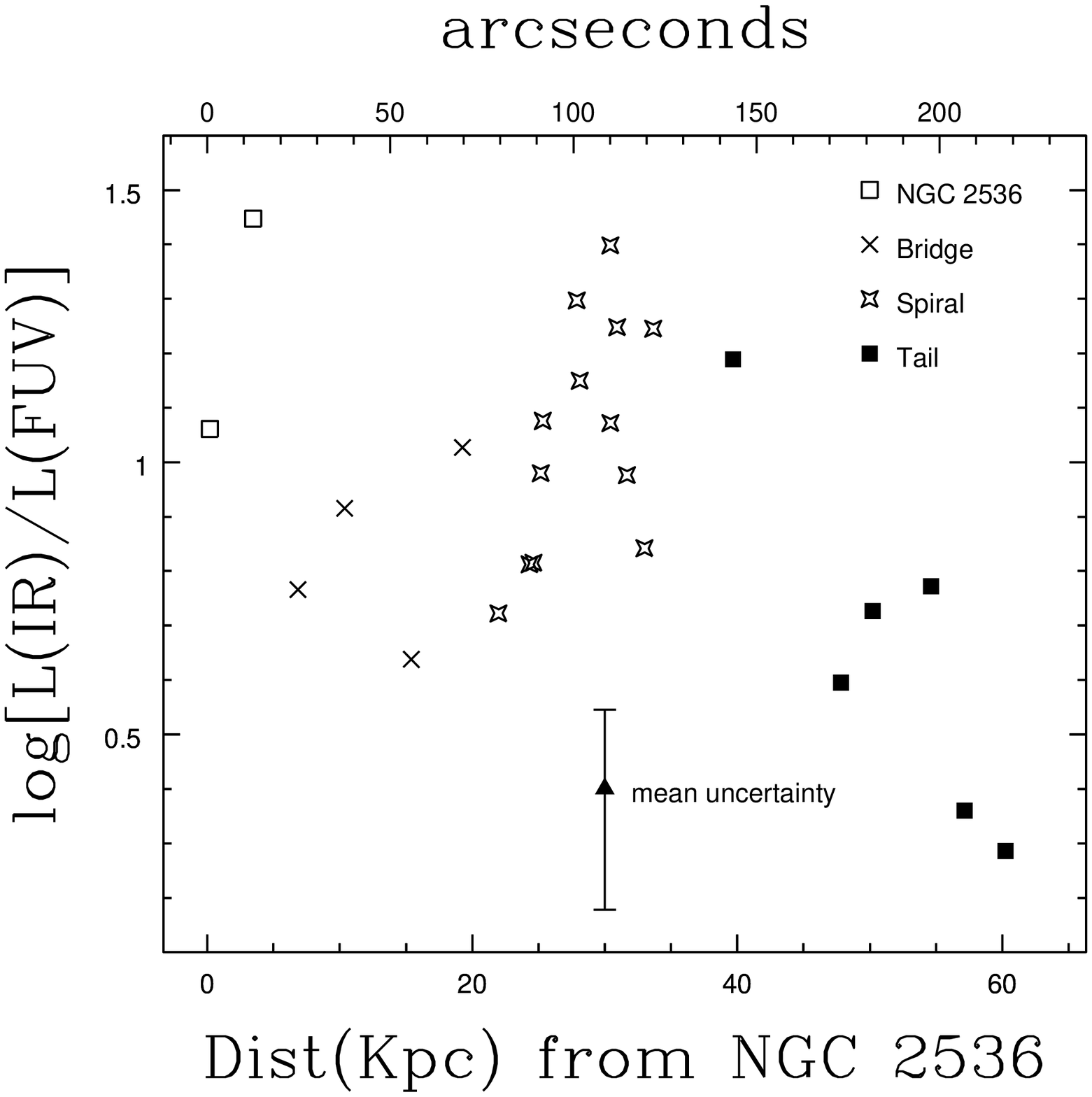}
\figcaption[f12.eps]{Log[L(IR)/L(FUV)] gradient.  The symbols are the same as in Figure 3. The mean uncertainty for the vertical axis reflects only the scatter in the L(IR) calibration.  \label{f12}}
\end{figure}

\newpage
\clearpage
\begin{figure}[ht]
\plotone{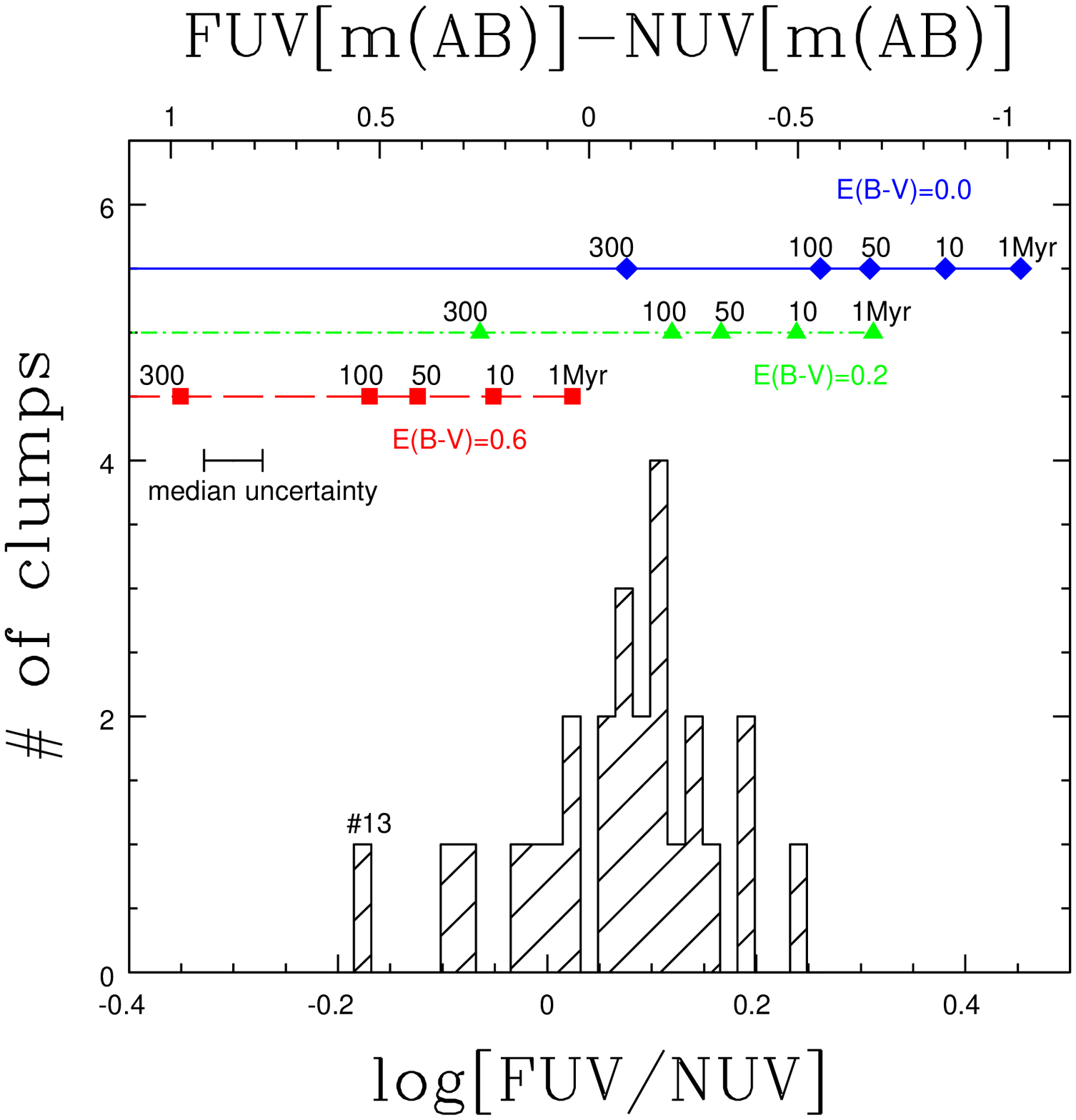}
\figcaption[f13.eps]{The log(FUV/NUV) distribution of the clumps in Arp 82.  At the top are SB99 models with solid line \ebv=0.0, dot-dash line \ebv=0.2 and dashed line \ebv=0.6. SB99 models generated assuming instantaneous star formation, solar abundances, and reddened with the \citet{cal94} reddening law.  Ages are marked with diamonds, triangles, and squares.  The median uncertainty in flux ratio is also shown.  The top axis is in magnitudes.\label{f13}}
\end{figure}

\begin{figure}[ht]
\plotone{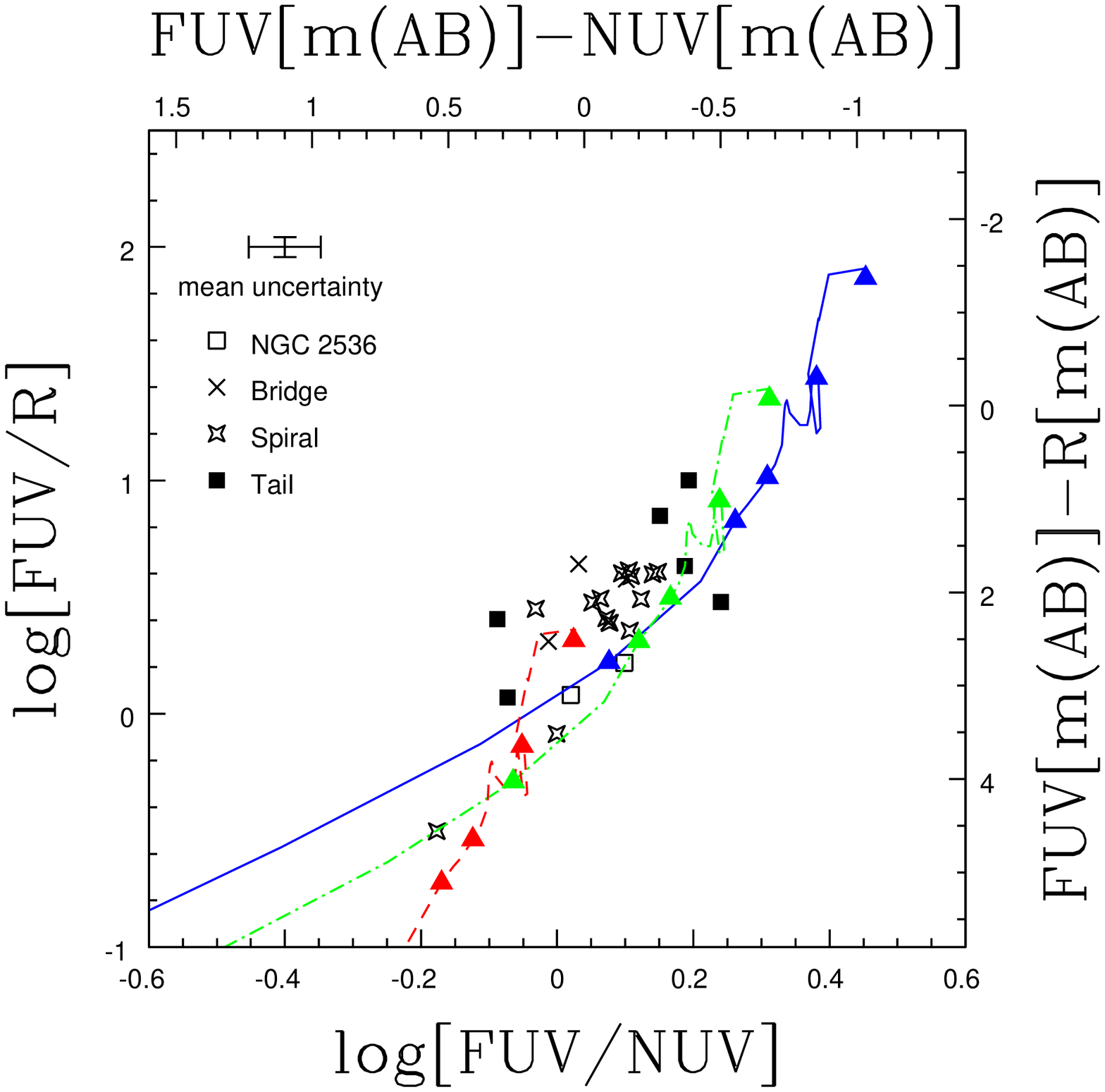}
\figcaption[f14.eps]{FUV/R vs FUV/NUV.  The symbols are the same as in Figure 3.  The SB99 models are: solid blue line and filled blue triangles \ebv=0.0, dot-dashed green line and filled green triangles \ebv=0.2 and dashed red line and filled red triangles \ebv=0.6. The SB99 models were generated assuming instantaneous star formation, solar abundances, and reddened with the \citet{cal94} reddening law.  The filled triangles plotted along the blue, green, and red curves represent ages of 1, 10, 50, 100, and 300 Myr respectively, with the youngest being at the upper right.  The median uncertainty in the clump's flux ratios is shown.  The top and right axes are magnitudes.  \label{f14}}
\end{figure}

\newpage
\clearpage
\begin{figure}[ht]
\plotone{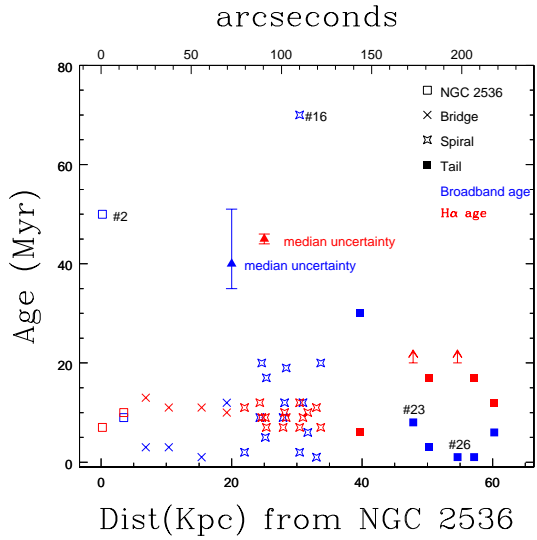}
\figcaption[f15.eps]{Clump age vs distance from NGC 2536.  The symbols are the same as in Figure 3.  Blue symbols are the broadband ages and red symbols are the ages determined from \ewha.  The uncertainties in ages reflect only the uncertainties in the measured fluxes. \label{f15}}
\end{figure}

\begin{figure}[ht]
\plotone{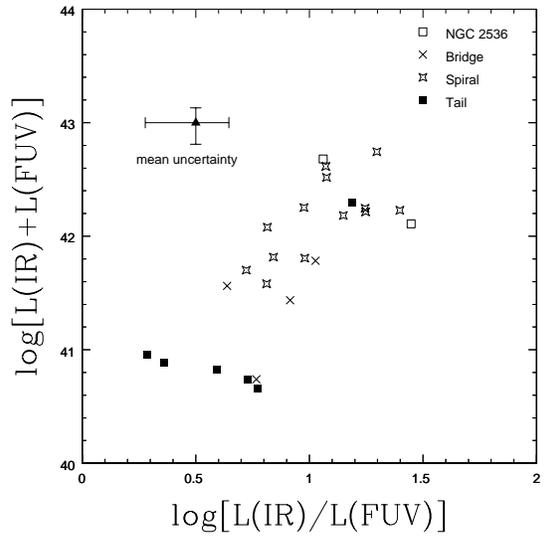}
\figcaption[f16.eps]{log[L(IR)+L(FUV)] vs log[L(IR)/L(FUV)].  The symbols are the same as in Figure 8.  The mean uncertainty reflects only the scatter in the L(IR) calibration. \label{f16}}
\end{figure}

\newpage
\clearpage
\begin{figure}[ht]
\plotone{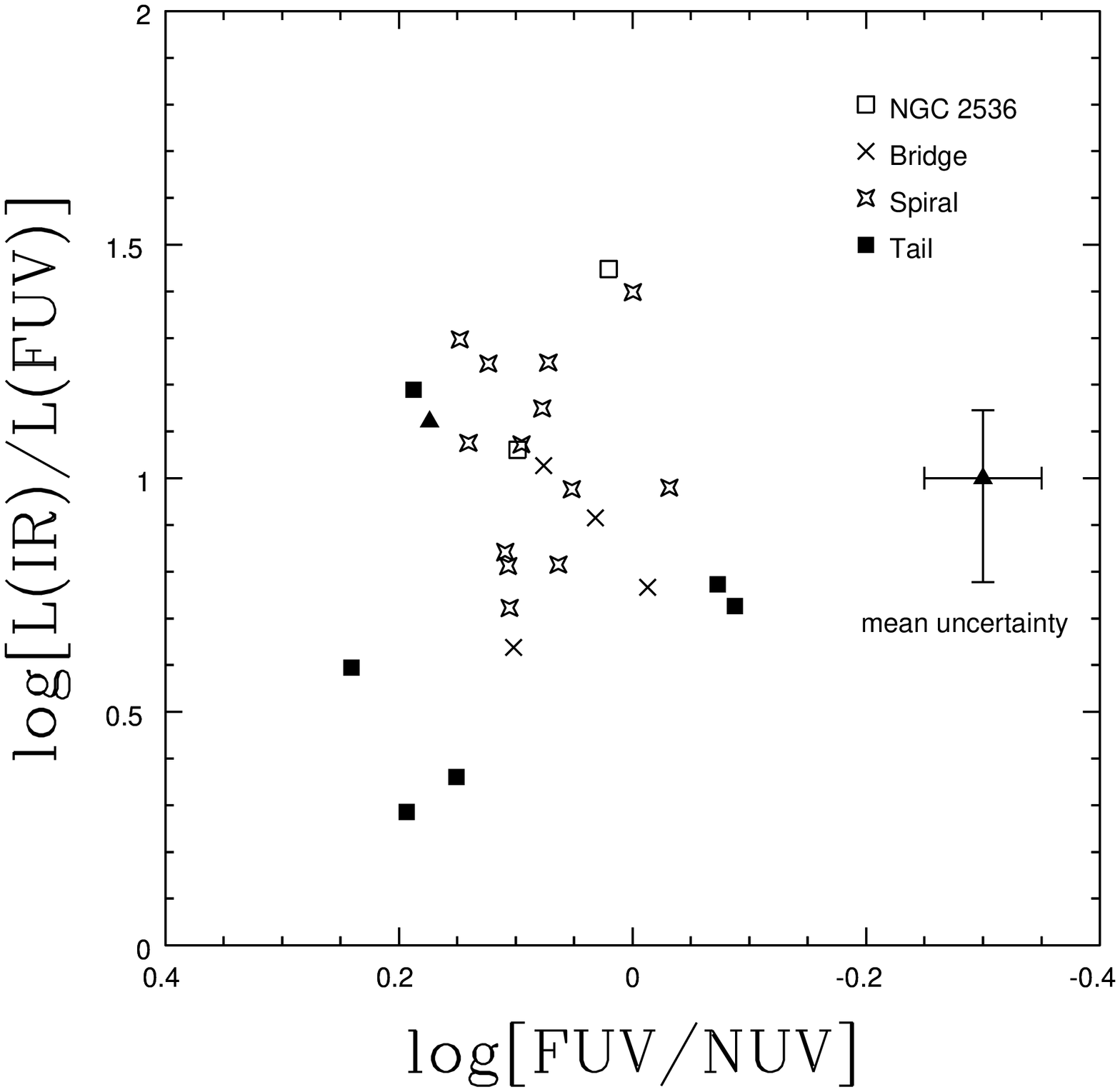}
\figcaption[f17.eps]{log[L(IR)/L(FUV)] vs log[FUV/NUV].  The symbols are the same as in Figure 8.  The mean uncertainty for the vertical axis reflects only the scatter in the L(IR) calibration.  \label{f17}}
\end{figure}

\begin{figure}[ht]
\plotone{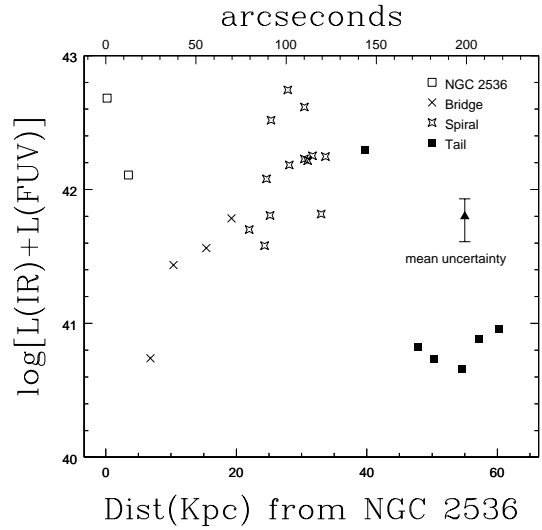}
\figcaption[f18.eps]{log[L(IR)+L(FUV)] vs distance from NGC 2536. The L(IR)+L(FUV) is a good proxy for the star formation rate.  The mean uncertainty reflects only the scatter in the L(IR) calibration.  \label{f18}}
\end{figure}

\newpage
\clearpage
\begin{figure}[ht]
\plotone{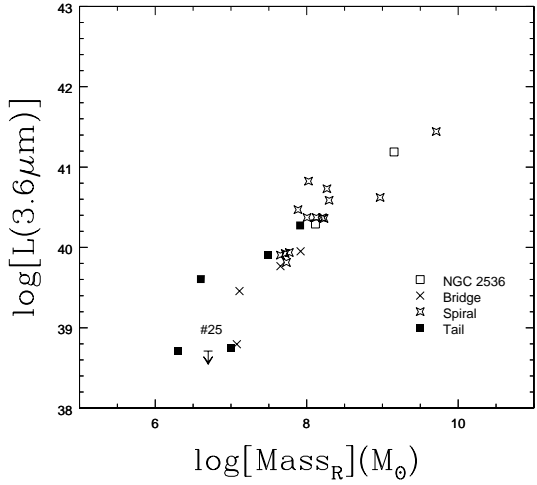}
\figcaption[f19.eps]{Mass$_{R}$ vs L(3.6\mic).  The mass is determined from SB99 and measured R band flux, see text.  The symbols are the same as in Figure 3. \label{f19}}
\end{figure}

\begin{figure}[ht]
\plotone{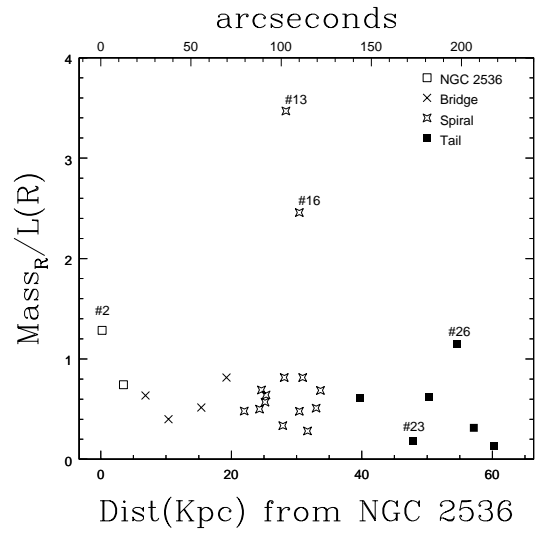}
\figcaption[f20.eps]{Mass to light ratio vs distance from NGC 2536.  The symbols are the same as in Figure 3.  \label{f20}}
\end{figure}

\newpage
\clearpage
\begin{figure}[ht]
\plotone{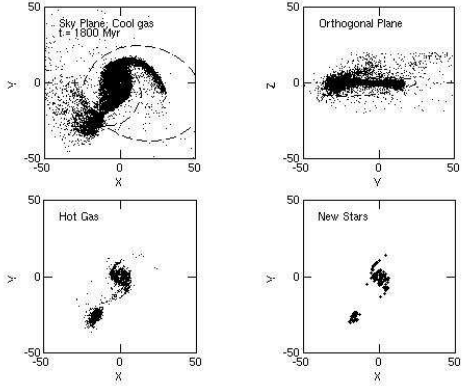} \figcaption[f21.eps]{Snapshots of the model at
a time near the present.  The upper panels show two orthogonal views
of the gas particles.  The primary's position is fixed at the
origin. The dashed curve also shows the orbit of the companion center
relative to the primary, starting from the point marked by a cross,
and continuing through the present to the future merger.  The
coordinate values are given in kpc with the adopted scaling. The lower
panels show gas particles that have either just turned on star
formation feedback (lower right), or have recently turned on feedback,
and so heated the gas (lower left). The latter thus gives a cumulative
picture of recent star formation, rather than just the immediate star
formation. \label{f21}}
\end{figure}

\begin{figure}[ht]
\plotone{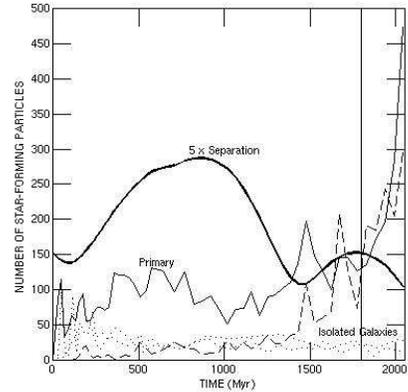}
\figcaption[f22.eps]{Shows star formation histories of the model
galaxies. Specifically, the thin solid curve shows the number of star
forming particles within 24 kpc of the center of the primary galaxy as
a function of time. The dashed curve shows the number of star forming
particles within 12 kpc of the center of the companion galaxy as a
function of time, and the dotted curves show the corresponding
quantities for the two galaxies in isolation. The values of the
integration radii were chosen to enclose the whole disk of the
individual galaxies, without including particles from the other
galaxy. The bold solid curve shows the separation between the galaxy
centers in kpc as a function of time, multiplied by a factor of 5 in
order to use the same y-axis scale. Comparison of this curve to the
thin solid and dashed curves allows one to see how the SFR depends on
separation.  The vertical line denotes the approximate present time. \label{f22}}
\end{figure}

\clearpage
\newpage

\begin{table}[ht]
\setlength{\tabcolsep}{0.06in}
\caption[Observation Log]{Observation Log}
\vspace{.2in}
\begin{tabular}{rlrcrc}
\hline \hline
\colzero  ID \colf Telescope\cola $\lambda_{central}$\colc Range \cold Exposure    \cole Date\eol
\colzero \colf        \cola   ($\mu$m)           \colc ($\mu$m)\cold ($\#\times$\ sec)      \cole (UT)\eol
\hline

\colzero 21197 \colf GALEX   \cola     0.153 \colc  0.135-0.175 \cold  $1\times1684$\cole  2005-02-24 \eol
\colzero 21197 \colf GALEX   \cola     0.231 \colc  0.175-0.280 \cold  $1\times3019$\cole  2005-02-13 \eol
\colzero 3247  \colf Spitzer \cola     3.6   \colc  3.0-4.2     \cold  $6\times12$  \cole 2004-11-01 \eol
\colzero 3247  \colf Spitzer \cola      4.5   \colc  3.7-5.3     \cold $6\times12$  \cole 2004-11-01 \eol
\colzero 3247  \colf Spitzer \cola    5.8   \colc  4.6-6.9     \cold  $6\times12$  \cole 2004-11-01 \eol
\colzero 3247  \colf Spitzer \cola     8.0   \colc  5.6-10.3    \cold $6\times12$ \cole 2004-11-01 \eol
\colzero 3247  \colf Spitzer \cola      24    \colc  18.0-32.2   \cold $32\times10$ \cole 2005-04-02 \eol
\colzero - - - \colf SARA   \cola      0.664 \colc  0.655-0.675 \cold  $16\times600$ \cole  2005-01-03  \eol
\colzero - - - \colf SARA   \cola      0.694 \colc  0.580-0.790 \cold  $5\times600$ \cole  2005-01-05  \eol

\hline\\
\end{tabular}
\end{table}

\clearpage
\newpage
\begin{table}[ht]
\setlength{\tabcolsep}{0.05in}
\caption[Clump Flux]{Clump Fluxes}
\vspace{.2in}
\begin{tabular}{rrrrrrrrrrrr}
\hline \hline

\colzero Clump\cola RA\colj  DEC\colk FUV$^1$\colb NUV$^1$\colc H$\alpha^1$\cold R$^1$\cole 3.6$^2$\colf 4.5$^2$\colg 5.8$^2$\colh 8.0$^2$\coli 24$^2$\eol
\colzero \cola (8:11)\colj  (+25)\colk \colb \colc \cold \cole $\mu$m\colf $\mu$m\colg $\mu$m\colh $\mu$m\coli $\mu$m\eol
\hline

\colzero     1\cola   15.061\colj  10:41.75\colk         2.43\colb            2.32\colc            1.58
\cold            2.02\cole            0.31\colf            0.23\colg            0.86
\colh            2.75\coli            4.94\eol
\colzero     2\cola  15.930\colj   10:46.47\colk         21.10\colb           16.82\colc            8.55
\cold           12.74\cole            2.45\colf            1.61\colg            3.69
\colh           10.27\coli           14.74\eol
\colzero     3\cola  14.452\colj  11:00.64\colk          0.44\colb            0.45\colc            0.25
\cold            0.22\cole            0.01\colf            $\leq$0.01\colg           $\leq$0.03
\colh            0.09\coli            0.27\eol
\colzero     4\cola  13.931\colj  11:12.44\colk          1.63\colb            1.51\colc            0.68
\cold            0.37\cole            0.05\colf            0.04\colg            0.19
\colh            0.52\coli            1.13\eol
\colzero     5\cola    12.887\colj  11:24.24\colk        3.76\colb            2.97\colc            0.93
\cold            1.00\cole            0.09\colf            0.07\colg            0.22
\colh            0.69\coli            0.99\eol
\colzero     6\cola   12.453\colj  11:38.40\colk         2.87\colb            2.41\colc            1.09
\cold            1.16\cole            0.14\colf            0.11\colg            0.42
\colh            1.31\coli            1.76\eol
\colzero     7\cola  11.757\colj  11:54.93\colk          2.80\colb            2.19\colc            0.50
\cold            1.23\cole            0.10\colf            0.07\colg            0.23
\colh            0.80\coli            0.92\eol
\colzero     8\cola  14.061\colj  12:02.60\colk          4.40\colb            3.45\colc            0.87
\cold            1.07\cole            0.13\colf            0.10\colg            0.30
\colh            0.94\coli            1.65\eol
\colzero     9\cola  12.627\colj  12:06.73\colk          3.34\colb            3.59\colc            1.54
\cold            1.18\cole            0.14\colf            0.11\colg            0.44
\colh            1.28\coli            2.37\eol
\colzero    10\cola   13.757\colj  12:11.45\colk         8.75\colb            7.56\colc            2.08
\cold            2.81\cole            0.36\colf            0.26\colg            0.81
\colh            2.30\coli            4.17\eol
\colzero    11\cola  11.931\colj  12:13.81\colk          5.54\colb            4.63\colc            1.39
\cold            2.26\cole            0.37\colf            0.24\colg            1.11
\colh            3.42\coli            4.16\eol
\colzero    12\cola  14.627\colj  12:17.35\colk         14.02\colb           10.15\colc            6.00
\cold            3.53\cole            0.61\colf            0.45\colg            1.91
\colh            5.98\coli           19.16\eol
\colzero    13\cola  13.409\colj  12:24.43\colk          5.33\colb            8.00\colc            6.67
\cold           16.96\cole            4.40\colf            3.03\colg            8.25
\colh           24.80\coli           76.04\eol
\colzero    14\cola  12.192\colj  12:25.61\colk         17.70\colb           14.23\colc            6.68
\cold            4.42\cole            0.85\colf            0.62\colg            2.92
\colh            8.56\coli           14.28\eol
\colzero    15\cola  15.149\colj  12:27.97\colk         14.63\colb           10.41\colc            5.91
\cold            3.61\cole            1.06\colf            0.79\colg            3.94
\colh           11.82\coli           20.26\eol
\colzero    16\cola  13.627\colj  12:33.28\colk          3.57\colb            3.57\colc            1.05
\cold            4.35\cole            0.66\colf            0.46\colg            1.12
\colh            3.24\coli            9.75\eol
\colzero    17\cola 12.714\colj  12:33.87\colk           9.38\colb            8.33\colc            1.64
\cold            3.12\cole            0.47\colf            0.33\colg            1.13
\colh            3.44\coli            7.56\eol
\colzero    18\cola  15.149\colj   12:39.18\colk          4.83\colb            4.09\colc            1.76
\cold            1.88\cole            0.37\colf            0.26\colg            1.18
\colh            3.58\coli            5.32\eol
\colzero    19\cola  14.018\colj  12:44.49\colk          4.53\colb            3.53\colc            0.75
\cold            1.16\cole            0.13\colf            0.10\colg            0.43
\colh            1.29\coli            2.15\eol
\colzero    20\cola  15.062\colj  12:49.21\colk          5.22\colb            3.93\colc            2.88
\cold            1.68\cole            0.37\colf            0.27\colg            1.28
\colh            3.86\coli            5.63\eol
\colzero    21\cola  13.931\colj  13:09.27\colk          6.61\colb            4.29\colc            7.11
\cold            1.54\cole            0.30\colf            0.26\colg            1.12
\colh            3.36\coli           16.19\eol
\colzero    22\cola  04.365\colj  13:22.25\colk          1.71\colb            1.09\colc            0.57
\cold            0.17\cole            0.01\colf            0.02\colg            0.05
\colh            0.11\coli            0.46\eol
\colzero    23\cola  11.582\colj  13:31.63\colk          0.74\colb            0.43\colc            $\leq$0.06
\cold            0.25\cole            0.06\colf            0.06\colg            0.07
\colh            0.09\coli            0.52\eol
\colzero    24\cola 10.282 \colj   13:33.88\colk           0.47\colb            0.58\colc            0.12
\cold            0.19\cole            0.01\colf            $\leq$0.01\colg            0.04
\colh            0.09\coli            0.30\eol
\colzero    25\cola  07.595\colj  13:43.09\colk          1.29\colb            0.91\colc            0.08
\cold            0.18\cole            $\leq$0.01\colf           $\leq$0.01\colg            $\leq$0.03
\colh            0.11\coli            0.29\eol
\colzero    26\cola  09.675\colj  13:47.97\colk          0.36\colb            0.43\colc            $\leq$0.06
\cold            0.31\cole            0.13\colf            0.09\colg            0.08
\colh            0.07\coli            0.43\eol
\hline
\colzero    Median\cola  \colj  \colk      \colb         \colc           \cold        \cole   \colf       \colg      \colh     \coli        \eol
\colzero    Uncert.\cola \colj  \colk       0.24\colb         0.15\colc           0.08\cold        0.02\cole   0.01\colf       0.01\colg      0.04\colh     0.04\coli        0.11\eol
\hline\\
\multicolumn{12}{l}{\footnotesize{$^1$ $10^{-16}$ erg s$^{-1}$ cm$^{-2}$ \AA$^{-1}$}}\\
\multicolumn{12}{l}{\footnotesize{$^2$ mJy; 1 mJy = $10^{-26}$ erg s$^{-1}$ cm$^{-2}$ Hz$^{-1}$}}\\
\end{tabular}
\end{table}

\clearpage 
\newpage
\begin{table}[ht]
\setlength{\tabcolsep}{0.05in}
\caption[Clump Colors]{Clump Colors}
\vspace{.2in}
\begin{tabular}{rrrrrrrrr}
\hline \hline

\colzero Clump\cola [3.6]\colb [3.6]-[4.5]\colc [4.5]-[5.8]\cold [5.8]-[8.0]\cole [8.0]-[24]\colf FUV\colg FUV-NUV\colh    FUV-R\eol

\colzero \cola (mag)$^1$\colb (mag)$^1$\colc (mag)$^1$\cold (mag)$^1$\cole (mag)$^1$\colf (mag)$^2$\colg (mag)$^2$\colh    (mag)$^{1,2}$\eol
\hline
\colzero     1\cola            14.9\colb             0.2\colc             1.9
\cold             1.9\cole             3.0\colf            20.7\colg             0.1
\colh             3.1\eol
\colzero     2\cola            12.7\colb             0.0\colc             1.4
\cold             1.7\cole             2.8\colf            18.4\colg            -0.1
\colh             2.8\eol
\colzero     3\cola            18.6\colb             $\leq$0.6\colc         - - -
\cold             $\geq$1.8\cole             3.5\colf            22.6\colg             0.1
\colh             2.5\eol
\colzero     4\cola            17.0\colb             0.4\colc             2.0
\cold             1.8\cole             3.2\colf            21.2\colg             0.0
\colh             1.7\eol
\colzero     5\cola            16.2\colb             0.2\colc             1.7
\cold             1.9\cole             2.8\colf            20.2\colg            -0.2
\colh             1.9\eol
\colzero     6\cola            15.7\colb             0.2\colc             1.9
\cold             1.9\cole             2.7\colf            20.5\colg            -0.1
\colh             2.3\eol
\colzero     7\cola            16.1\colb             0.1\colc             1.7
\cold             2.0\cole             2.5\colf            20.6\colg            -0.2
\colh             2.4\eol
\colzero     8\cola            15.9\colb             0.2\colc             1.7
\cold             1.9\cole             3.0\colf            20.1\colg            -0.2
\colh             1.8\eol
\colzero     9\cola            15.8\colb             0.2\colc             2.0
\cold             1.8\cole             3.0\colf            20.4\colg             0.2
\colh             2.2\eol
\colzero    10\cola            14.7\colb             0.1\colc             1.7
\cold             1.8\cole             3.0\colf            19.3\colg            -0.1
\colh             2.1\eol
\colzero    11\cola            14.7\colb             0.0\colc             2.1
\cold             1.9\cole             2.6\colf            19.8\colg            -0.1
\colh             2.3\eol
\colzero    12\cola            14.2\colb             0.1\colc             2.1
\cold             1.9\cole             3.6\colf            18.8\colg            -0.3
\colh             1.8\eol
\colzero    13\cola            12.0\colb             0.1\colc             1.6
\cold             1.8\cole             3.6\colf            19.9\colg             0.5
\colh             4.6\eol
\colzero    14\cola            13.8\colb             0.1\colc             2.2
\cold             1.8\cole             2.9\colf            18.6\colg            -0.1
\colh             1.8\eol
\colzero    15\cola            13.6\colb             0.2\colc             2.2
\cold             1.8\cole             2.9\colf            18.8\colg            -0.3
\colh             1.8\eol
\colzero    16\cola            14.1\colb             0.1\colc             1.5
\cold             1.8\cole             3.6\colf            20.3\colg             0.1
\colh             3.5\eol
\colzero    17\cola            14.4\colb             0.1\colc             1.8
\cold             1.8\cole             3.2\colf            19.3\colg            -0.0
\colh             2.1\eol
\colzero    18\cola            14.7\colb             0.1\colc             2.1
\cold             1.8\cole             2.8\colf            20.0\colg            -0.1
\colh             2.3\eol
\colzero    19\cola            15.8\colb             0.2\colc             2.0
\cold             1.8\cole             2.9\colf            20.0\colg            -0.2
\colh             1.8\eol
\colzero    20\cola            14.7\colb             0.2\colc             2.2
\cold             1.8\cole             2.8\colf            19.9\colg            -0.2
\colh             2.1\eol
\colzero    21\cola            14.9\colb             0.3\colc             2.1
\cold             1.8\cole             4.1\colf            19.6\colg            -0.4
\colh             1.7\eol
\colzero    22\cola            18.8\colb             1.2\colc             1.7
\cold             1.5\cole             3.9\colf            21.1\colg            -0.4
\colh             0.8\eol
\colzero    23\cola            16.6\colb             0.4\colc             0.7
\cold             0.9\cole             4.2\colf            22.0\colg            -0.5
\colh             2.1\eol
\colzero    24\cola            18.8\colb             $\leq$0.8\colc             $\geq$1.9
\cold             1.4\cole             3.7\colf            22.5\colg             0.3
\colh             2.3\eol
\colzero    25\cola            $\geq$18.8\colb            - - -\colc            - - -
\cold             $\geq$2.0\cole             3.4\colf            21.4\colg            -0.3
\colh             1.2\eol
\colzero    26\cola            15.8\colb             0.2\colc             0.2
\cold             0.5\cole             4.4\colf            22.8\colg             0.3
\colh             3.1\eol

\hline
\colzero    Median\cola        \colb         \colc           \cold        \cole   \colf       \colg    \colh    \eol
\colzero    Uncert.\cola     0.1\colb   0.1\colc  0.2\cold    0.1\cole   0.1\colf   0.1\colg  0.1 \colh   0.1 \eol

\hline\\
\multicolumn{9}{l}{\footnotesize{$^1$ magnitudes in AB system \citet{oke74}}}\\
\multicolumn{9}{l}{\footnotesize{$^2$ magnitudes in AB system \citet{oke90}}}\\
\end{tabular}
\end{table}

\clearpage
\newpage
\begin{table}[ht]
\caption[Ages and E(B$-$V)s]{Ages and E(B$-$V)s}
\vspace{.2in}
\begin{tabular}{rcrrrr}
\hline \hline

\colzero Clump\cola  Region\colf Age\colb Age\colc E(B$-$V)\cold E(B$-$V)\eol
\colzero \cola  \colf [Broadband]\colb [EW(H$\alpha$)]\colc [Broadband]\cold [L(IR)]\eol
\colzero \cola \colf (Myr)\colb (Myr)\colc (mag)\cold (mag)\eol
\hline
\colzero     1\cola NGC2536\colb               9$^{+191}_{-7}$\colc              10$^{+0}_{-1}$
\cold             0.5$^{+0.2}_{-0.3}$\cole             0.9$\pm0.1$\eol
\colzero     2\cola NGC2536\colb              50$^{+10}_{-36}$\colc               7$^{+0}_{-0}$
\cold             0.3$^{+0.1}_{-0.0}$\cole             0.6$\pm0.1$\eol
\colzero     3\cola bridge\colb               3$^{+5}_{-0}$\colc              13$^{+0}_{-0}$
\cold             0.6$^{+0.1}_{-0.3}$\cole             0.5$\pm0.1$\eol
\colzero     4\cola bridge\colb               3$^{+5}_{-0}$\colc              11$^{+0}_{-1}$
\cold             0.5$^{+0.0}_{-0.1}$\cole             0.6$\pm0.1$\eol
\colzero     5\cola bridge\colb               1$^{+17}_{-0}$\colc              11$^{+0}_{-1}$
\cold             0.5$^{+0.0}_{-0.2}$\cole             0.4$\pm0.1$\eol
\colzero     6\cola bridge\colb              12$^{+18}_{-10}$\colc              10$^{+0}_{-0}$
\cold             0.4$^{+0.2}_{-0.1}$\cole             0.6$\pm0.1$\eol
\colzero     7\cola spiral\colb               9$^{+191}_{-0}$\colc              12$^{+1}_{-0}$
\cold             0.4$^{+0.0}_{-0.3}$\cole             0.5$\pm0.1$\eol
\colzero     8\cola spiral\colb               2$^{+15}_{-1}$\colc              11$^{+0}_{-0}$
\cold             0.5$^{+0.0}_{-0.2}$\cole             0.5$\pm0.1$\eol
\colzero     9\cola spiral\colb               5$^{+0}_{-2}$\colc               9$^{+1}_{-0}$
\cold             0.5$^{+0.1}_{-0.0}$\cole             0.6$\pm0.1$\eol
\colzero    10\cola spiral\colb              20$^{+0}_{-14}$\colc               9$^{+1}_{-1}$
\cold             0.3$^{+0.1}_{-0.0}$\cole             0.5$\pm0.1$\eol
\colzero    11\cola spiral\colb              12$^{+0}_{-10}$\colc              10$^{+0}_{-0}$
\cold             0.4$^{+0.2}_{-0.0}$\cole             0.7$\pm0.1$\eol
\colzero    12\cola spiral\colb              17$^{+0}_{-12}$\colc               7$^{+0}_{-0}$
\cold             0.3$^{+0.2}_{-0.0}$\cole             0.7$\pm0.1$\eol
\colzero    13\cola spiral\colb              19$^{+1}_{-18}$\colc               9$^{+0}_{-1}$
\cold             0.7$^{+0.2}_{-0.0}$\cole             1.2$\pm0.1$\eol
\colzero    14\cola spiral\colb               2$^{+0}_{-1}$\colc               7$^{+0}_{-0}$
\cold             0.5$^{+0.0}_{-0.0}$\cole             0.7$\pm0.1$\eol
\colzero    15\cola spiral\colb               9$^{+8}_{-7}$\colc               7$^{+0}_{-0}$
\cold             0.3$^{+0.2}_{-0.0}$\cole             0.8$\pm0.1$\eol
\colzero    16\cola spiral\colb              70$^{+0}_{-63}$\colc              12$^{+0}_{-0}$
\cold             0.4$^{+0.2}_{-0.0}$\cole             0.8$\pm0.1$\eol
\colzero    17\cola spiral\colb               6$^{+14}_{-1}$\colc              10$^{+1}_{-0}$
\cold             0.4$^{+0.1}_{-0.1}$\cole             0.6$\pm0.1$\eol
\colzero    18\cola spiral\colb              12$^{+0}_{-8}$\colc               9$^{+1}_{-0}$
\cold             0.4$^{+0.1}_{-0.0}$\cole             0.7$\pm0.1$\eol
\colzero    19\cola spiral\colb               1$^{+16}_{-0}$\colc              11$^{+0}_{-0}$
\cold             0.5$^{+0.0}_{-0.2}$\cole             0.5$\pm0.1$\eol
\colzero    20\cola spiral\colb              20$^{+0}_{-7}$\colc               7$^{+0}_{-0}$
\cold             0.3$^{+0.0}_{-0.0}$\cole             0.7$\pm0.1$\eol
\colzero    21\cola tail\colb              30$^{+60}_{-20}$\colc               6$^{+0}_{-0}$
\cold             0.2$^{+0.1}_{-0.1}$\cole             0.7$\pm0.1$\eol
\colzero    22\cola tail\colb               6$^{+63}_{-1}$\colc              12$^{+1}_{-1}$
\cold             0.2$^{+0.1}_{-0.2}$\cole             0.3$\pm0.1$\eol
\colzero    23\cola tail?\colb               8$^{+192}_{-7}$\colc              20$^{+2}_{-0}$
\cold             0.3$^{+0.2}_{-0.3}$\cole             0.4$\pm0.1$\eol
\colzero    24\cola tail\colb               3$^{+16}_{-2}$\colc              17$^{+1}_{-2}$
\cold             0.6$^{+0.0}_{-0.2}$\cole             0.5$\pm0.1$\eol
\colzero    25\cola tail\colb               1$^{+12}_{-0}$\colc              17$^{+1}_{-1}$
\cold             0.4$^{+0.0}_{-0.2}$\cole             0.3$\pm0.1$\eol
\colzero    26\cola tail?\colb               1$^{+49}_{-0}$\colc              20$^{+3}_{-0}$
\cold             0.7$^{+0.1}_{-0.4}$\cole             0.5$\pm0.1$\eol

\hline\\
\end{tabular}
\end{table}
  
\clearpage
\newpage
\begin{table}[ht]
\caption[Clump Masses]{Clump Masses}
\vspace{.2in}
\begin{tabular}{rcrr}
\hline \hline
\colzero Clump\cola Region\colb Mass$_{R}^1$\colc Mass$_{UV}^2$\eol
\colzero   \cola   \colb  ($10^6$M$_\odot$)\colc ($10^6$M$_\odot$)\eol
\hline
\colzero     1\cola NGC2536\colb      131$^{+2839}_{-94}$\colc      124$^{+61776}_{-121}$\eol
\colzero     2\cola NGC2536\colb     1430$^{+1020}_{-959}$\colc     1370$^{+3350}_{-1098}$\eol
\colzero     3\cola bridge\colb       12$^{+5}_{-9}$\colc       11$^{+28}_{10.5}$\eol
\colzero     4\cola bridge\colb       13$^{+0}_{-4}$\colc       15$^{+54}_{-10}$\eol
\colzero     5\cola bridge\colb       45$^{+87}_{-25}$\colc       44$^{+411}_{-38}$\eol
\colzero     6\cola bridge\colb       83$^{+231}_{-62}$\colc       82$^{+1878}_{-78}$\eol
\colzero     7\cola spiral\colb       54$^{+488}_{-38}$\colc       52$^{+3508}_{-49}$\eol
\colzero     8\cola spiral\colb       45$^{+90}_{-24}$\colc       44$^{+449}_{-37}$\eol
\colzero     9\cola spiral\colb       59$^{+29}_{-16}$\colc       65$^{+111}_{-35}$\eol
\colzero    10\cola spiral\colb      170$^{+85}_{-124}$\colc      169$^{+291}_{-139}$\eol
\colzero    11\cola spiral\colb      161$^{+195}_{-99}$\colc      158$^{+1022}_{-138}$\eol
\colzero    12\cola spiral\colb      198$^{+258}_{-121}$\colc      211$^{+1369}_{-175}$\eol
\colzero    13\cola spiral\colb     5140$^{+6460}_{-3440}$\colc     5220$^{+36380}_{-4754}$\eol
\colzero    14\cola spiral\colb      185$^{+11}_{-1}$\colc      177$^{+32}_{-0}$\eol
\colzero    15\cola spiral\colb      106$^{+351}_{-40}$\colc      100$^{+1540}_{-80}$\eol
\colzero    16\cola spiral\colb      934$^{+1166}_{-855}$\colc      974$^{+6226}_{-930}$\eol
\colzero    17\cola spiral\colb       77$^{+348}_{-9}$\colc       87$^{+1253}_{-63}$\eol
\colzero    18\cola spiral\colb      134$^{+65}_{-89}$\colc      138$^{+238}_{-113}$\eol
\colzero    19\cola spiral\colb       52$^{+95}_{-29}$\colc       54$^{+454}_{-47}$\eol
\colzero    20\cola spiral\colb      101$^{+0}_{-37}$\colc      101$^{+0}_{-40}$\eol
\colzero    21\cola tail\colb       82$^{+181}_{-57}$\colc       82$^{+211}_{-75}$\eol
\colzero    22\cola tail\colb        2$^{+22}_{-1}$\colc        2$^{+169}_{-1.8}$\eol
\colzero    23\cola tail?\colb        4$^{+158}_{-3}$\colc        4$^{+2556}_{-3.9}$\eol
\colzero    24\cola tail\colb       10$^{+27}_{-5}$\colc       12$^{+157}_{-10}$\eol
\colzero    25\cola tail\colb        5$^{+5}_{-3}$\colc        6$^{+35}_{-5}$\eol
\colzero    26\cola tail?\colb       31$^{+231}_{-25}$\colc       32$^{+3498}_{-31}$\eol


\hline\\
\multicolumn{4}{l}{\footnotesize{$^1$ Mass determined from R band flux and SB99.  (See text)}}\\
\multicolumn{4}{l}{\footnotesize{$^2$ Mass determined from FUV band flux and SB99.  (See text)}}\\
\end{tabular}
\end{table}

\end{document}